\documentclass[a4paper,11pt]{article}
\pdfoutput=1 
\usepackage{verbatim}

\usepackage{jheppub}
\usepackage{amsmath,amsfonts,amssymb,amsthm,graphicx,color}
\usepackage[T1]{fontenc} 
\usepackage{enumerate}
\usepackage{tikz, tikz-3dplot, pgfplots}
\usepackage[utf8]{inputenc}
\usepackage{slashed}
\usepackage{subcaption}

\usepackage{changepage}
\usepackage{float}
\usepackage{tikz}
\usetikzlibrary{decorations.text}
\usepackage{enumitem}
\setlist{itemsep=0pt}
\usetikzlibrary{math}

\newcommand{\fig}{Fig.~\ref}

\usepackage{tikz}
\usepackage{pgfplots}
\usetikzlibrary{decorations.text,intersections, pgfplots.fillbetween}
\usetikzlibrary{math}
 \usetikzlibrary{snakes,3d,shapes.geometric,shadows.blur}
\usepackage{tikz-3dplot}

\makeatletter
\newcommand\footnoteref[1]{\protected@xdef\@thefnmark{\ref{#1}}\@footnotemark}
\makeatother

\newcommand{\comm}[1]{} 

\def\({\left(}
\def\){\right)}
\def\[{\left[}
\def\]{\right]}

\def\One{{\hbox{ 1\kern-.8mm l}}}

\def\barray{\begin{array}}
\def\earray{\end{array}}
\def\be{\begin{equation}}
\def\ee{\end{equation}}
\def\bea{\begin{eqnarray}}
\def\eea{\end{eqnarray}}
\def\bal{\begin{align}}
\def\eal{\end{align}}

\def\-{\,-\,}
\def\={\,=\,}
\def\+{\,+\,}
\def\equi{\,\equiv\,}

\numberwithin{equation}{section} 

\definecolor{cardinal}{rgb}{0.6,0,0}
\definecolor{darkgreen}{rgb}{0,0.4,0}
\definecolor{golden}{rgb}{0.92, 0.7, 0}
\definecolor{midnight}{rgb}{0, 0, 0.5}
\definecolor{darkblue}{rgb}{0, 0, 0.7}
\definecolor{purple}{rgb}{0.5, 0, 0.5}

\def\IR{\mathbb{R}}

\def\cA{{\cal A}}

\def\cG{{\cal G}}

\def\cL{{\cal L}}
\def\cM{{\cal M}}
\def\cN{{\cal N}}

\def\cO{{\cal O}}
\def\cQ{{\cal Q}}

\def\cZ{{\cal Z}}

\usetikzlibrary{positioning,calc,fadings,decorations.pathreplacing}

\tikzset{
 diffuse color/.initial = black,                       
}

\tikzset{
 linear opacity/.initial=0.5,                          
 linear stroke/.style = {                              
   preaction={                                         
     draw=\pgfkeysvalueof{/tikz/diffuse color},        
     line width = (2.0-#1)*\pgflinewidth,              
     opacity=\pgfkeysvalueof{/tikz/linear opacity},white}},  
 diffuse gradient/.style={                             
   draw = none,                                        
   linear opacity=#1,                                  
   linear stroke/.list={0.0,#1,...,1.0}},              
 diffuse gradient/.default=1,                          
}

\tikzset{
 non-linear stroke/.style = {                          
   preaction={                                         
     draw=\pgfkeysvalueof{/tikz/diffuse color},        
     line width = (2.0-#1)*\pgflinewidth,              
     opacity=#1,white}},                                     
 diffuse falloff/.style={                              
   draw = none,                                        
   non-linear stroke/.list={0.0,#1,...,1.0}},          
 diffuse falloff/.default=1,                           
}
 
\tikzset{%
  >=latex, 
  inner sep=0pt,%
  outer sep=2pt,%
  mark coordinate/.style={inner sep=0pt,outer sep=0pt,minimum size=3pt,
    fill=black,circle}%
}

\title{\boldmath Bubble Bag End: A Bubbly Resolution of Curvature Singularity
}

\author{Ibrahima Bah and}
\author{Pierre Heidmann} 
\affiliation{Department of Physics and Astronomy, Johns Hopkins University, 3400 North Charles Street, Baltimore, MD 21218, USA}

\emailAdd{iboubah@jhu.edu}
\emailAdd{pheidma1@jhu.edu}

\abstract{We construct a family of smooth charged bubbling solitons in $\mathbb{M}^4 \times$T$^2$, four-dimensional Minkowski with a two-torus.  The solitons are characterized by a degeneration pattern of the torus along a line in $\mathbb{M}^4$ defining a chain of topological cycles. They live in the same parameter regime as non-BPS non-extremal four-dimensional black holes, and are ultra-compact with sizes ranging from miscroscopic to macroscopic scales. The six-dimensional framework can be embedded in type IIB supergravity where the solitons are identified with geometric transitions of non-BPS D1-D5-KKm bound states.  Interestingly, the geometries admit a minimal surface that smoothly opens up to a bubbly end of space.  Away from the solitons, the solutions are indistinguishable from a new class of singular geometries.  By taking a limit of large number of bubbles, the soliton geometries can be matched arbitrarily close to the singular spacetimes.  This provides the first classical resolution of a curvature singularity beyond the framework of supersymmetry and supergravity by blowing up topological cycles wrapped by fluxes at the vicinity of the singularity.
}

\preprint{}
\begin{document}

\maketitle
\flushbottom

\newpage

\section{Introduction}
\label{sec:Intro}

The resolution of singularities in general relativity (GR) is of paramount interest as it holds the key to uncovering fundamental aspects of black holes and big bang initial conditions.  In general, singularities in physical theories signal presence of new degrees, which in the case of black holes or cosmological singularities, are expected to generically correspond to new states of quantum gravity.  On the other hand, quantum states can be coherent and admit descriptions in classical theories.  Particularly in classical theories of gravity, it is interesting to ask whether there exist such coherent states that resolve singularities.  These would correspond to solitons of the nonlinear field equations, i.e. smooth asymptotically flat stationary solutions with finite energy.  

The question on the existence of solitons in gravity is as old as the subject, dating back to the work of Serini in 1918 \cite{SeriniSoliton}.  This was pushed forward by Einstein \cite{EinsteinSoliton}, Einstein-Pauli \cite{Einstein:1943ixi} and Lichnerowicz \cite{LichSoliton}.  The results in these work lead to the theorem in four dimensions: {\it If a  solution is  asymptotically flat, 
topologically  trivial  and globally  stationary  (i.e. admits an  everywhere  timelike 
Killing vector field)  then it must be flat  space}.  This is a no-go theorem for solitons that are topologically trivial.\footnote{For an enlightening discussion of the interesting history of solitons in gravity, see \cite{Gibbons:2013tqa,Breitenlohner:1987dg} and references thereof.}  

More recently, Gibbons and Warner \cite{Gibbons:2013tqa} revisited the soliton no-go theorem in the context of supergravity and distilled important loopholes for evading them.  These involve allowing for interesting topological structures in spacetime, topological interactions such Chern-Simons terms and nontrivial electromagnetic fluxes to support them.  A particular interest of this work is to understand how microstate geometries \cite{Mathur:2005zp,Bena:2013dka} -- smooth horizonless geometries that describe coherent black hole microstates -- could exist in theories of gravity.  

There is an extensive body of literature \cite{Bena:2004de,Mathur:2005zp,Bena:2006kb,Bena:2007qc,Bena:2007kg,Bena:2013dka,Bena:2015bea,Heidmann:2017cxt,Bena:2017fvm,Heidmann:2019xrd} describing supersymmetric solitonic solutions that resolve black hole horizons.  In this paper, we are interested in describing non-supersymmetric solitons in general relativity that resolve curvature singularities.  We employ a novel mechanism discovered by the authors in \cite{Bah:2020ogh,Bah:2020pdz,ourpaper}, motivated by \cite{Stotyn:2011tv,Elvang:2002br,Gibbons:2013tqa}, to construct solitons with interesting topology in the six dimensional space $\mathbb{M}^4\times $T$^2$ (four dimensional Minkowski space times a two-torus) without the help of supersymmetry. 

The work in \cite{ourpaper} provides a new mechanism that allows for {\it linear superposition} of smooth, regular and charged solitonic objects in gravity beyond supersymmetry.\footnote{Previous work \cite{Jejjala:2005yu,Bossard:2014ola,Bena:2015drs,Bena:2016dbw,Bossard:2017vii} has extended the scope of the microstate geometry program by developing a mechanism for constructing non-supersymmetric charged solitons that resolve the horizon of non-extremal unphysical black holes in five dimensions. Their mechanism is different from ours, resulting in very different solutions. However, their method is also to linearize Einstein equations without the help of supersymmetry while retaining sufficient degrees of freedom for the construction of non-trivial topological structures.}  The scope of the result is unprecedented given the non-linear structure of Einstein equations.  The constructions provide a framework for describing and studying new classes of ultra-compact objects with arbitrary size from microscopic to macroscopic scales.  As interestingly, these solutions are completely characterized within the gravitational theory without any appeal to new UV physics.  On the other hand, they naturally embed in string theory where the topological structure can be reinterpreted as geometric transition of strings and branes.  Our mechanism also provides the first resolution of curvature singularities using topology in gravitational theories without the aid of supersymmetry.  

Our new ultra-compact objects can live in the same regime of parameter space as non-supersymmetric and non-extremal black holes, and therefore have particular importance for astrophysical studies and observation of black holes by gravitational-wave detectors.  The phenomenological properties of our constructions can provide a new way to test gravity and its possible higher-dimensional nature.  In particular, these topological structures can correspond to new \emph{geometric phases of matter}.  

\subsubsection*{New mechanism for bubbling geometries}

The construction of solitonic objects in theories of gravity with extra compact dimensions by directly treating Einstein equations has been historically a long and winding road. Starting with five dimensions, i.e. a $\mathbb{M}^4\times$S$^1$ background (four dimensional Minkowski times a circle),  there exist spherically symmetric static {\it vacuum} solutions with a locus where the S$^1$ collapses to zero size.  In this region, the spacetime is smooth and decomposes as $\mathbb{R}_t \times $S$^2 \times \mathbb{R}^2$ -- the time direction, a finite size S$^2$ and an $\mathbb{R}^2$ obtained from a radial direction combined with the collapsing circle.  The spacetime is smooth and carries zero energy in five dimensions.  It is topologically non-trivial as it admits a finite size S$^2$, referred to as a "bubble of nothing" in its interior.  This object in vacuum is unstable and wants to grow, and in doing so mediate a decay of the Kaluza-Klein vacuum \cite{Witten:1981gj}.  

The instability of the bubble of nothing can be mitigated by considering Kaluza-Klein (KK) backgrounds with non-trivial electromagnetic fluxes wrapped on the S$^2$.  The presence of the fluxes lifts the vacuum to a background with a non-trivial ADM mass and electromagnetic charges; more importantly it provides a counterbalance to the expanding bubble and stabilizes it to a finite size \cite{Stotyn:2011tv,Bah:2020ogh,Bah:2020pdz}.\footnote{Even in the stable case, regularity conditions of the simple smooth bubble leads to a bound of its diameter to be smaller than the radius of the KK circle.  An important result in \cite{Bah:2020ogh,Bah:2020pdz} is that the regularity conditions can be loosen by allowing orbifold singularities which lifts the bound on the size of the bubble.  These orbifold singularities can then be resolved with Gibbons-Hawkins bubbles.  This leads to the proposal of new astrophysically interesting ultra-compact objects dubbed topological stars \cite{Bah:2020ogh,Bah:2020pdz}.}

\begin{figure}[h]\centering
\begin{subfigure}{0.46\textwidth}\centering
 \includegraphics[width=\textwidth]{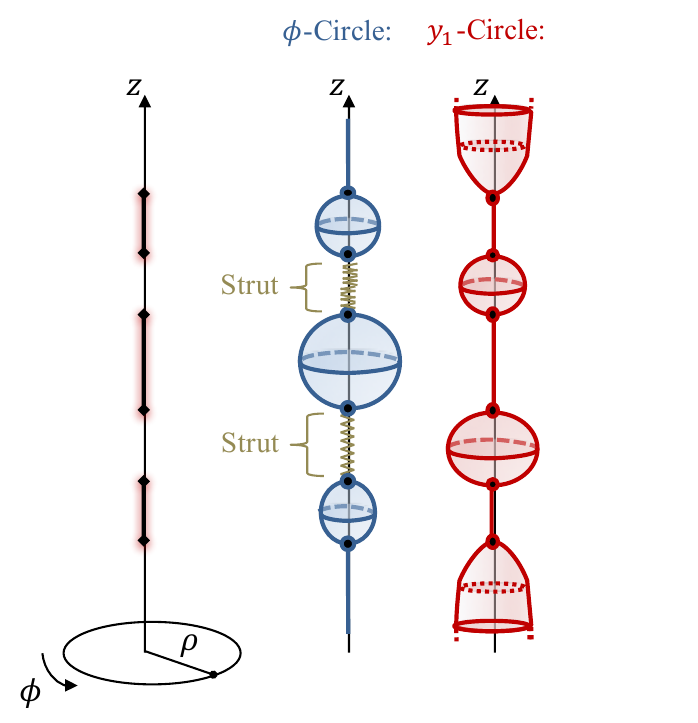}
 \caption{Bubble-strut system in five dimensions}\label{bubblestrut}
 \end{subfigure}
 \hspace{0.5cm}
 \begin{subfigure}{0.46\textwidth}\centering
   \includegraphics[width=\textwidth]{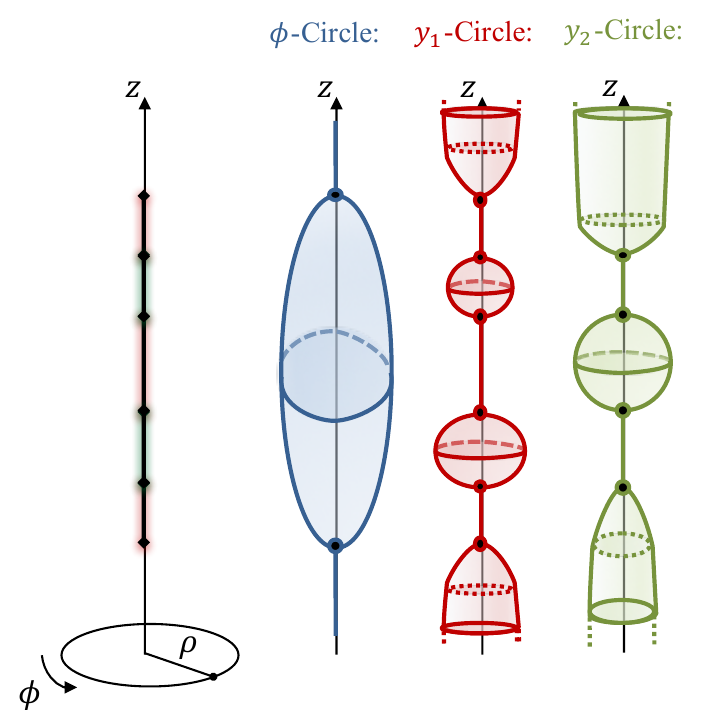}
    \caption{Struts replaced with new species of rods}\label{multibubble}
 \end{subfigure}
 \caption{In \fig{bubblestrut}, there are multiple rod sources along the $z$-axis of three dimensional cylindrical base.  Each rod source corresponds to a region where the KK circle with coordinate $y_1$ collapses defining a bubble locus.  The struts are segments where the base circle shrinks.  Solutions with arbitrary chain of struts and bubbles can be constructed.  In \fig{multibubble}, we add a new circle and substitute the struts with a new species of rod sources.  This leads to a smooth topological structure that grows with the number of rod sources.}
 \label{fig:5dWeylIntro}
\end{figure}

The charged bubble is a non-trivial solution to the non-linear equations of gravity.  A more careful study on how they solve the equations, allowed us in \cite{Bah:2020pdz} to discover a framework to add electromagnetic charges and fluxes to vacuum Weyl solutions of \cite{Weyl:book,Emparan:2001wk}.  Here, a family of static multi-bubble solutions with nontrivial electromagnetic charges that are supported along a line with struts between them has been constructed.  In \fig{bubblestrut}, we summarize the structure of the solutions.  The rod sources correspond to regions where the KK S$^1$, with coordinate $y_1$, shrinks forming a finite size charged bubble. The struts are strings of negative tension that correspond to regions where the $\phi$-circle that parametrizes the $U(1)$ action around the structure in $\mathbb{M}^4$ shrinks.  The struts provide a counterbalance to the attraction of the bubbles.  In the limit where the strut shrinks to zero size, two neighboring bubbles merge to a single bubble. Throughout the paper, we use the term rod source and bubble synonymously.   

Although the charged multi-bubble solutions of \cite{Bah:2020pdz} possess no curvature singularities or horizons, they contain struts which are regions with negative energy density with no known UV origin.  An important question is whether topology can be used in lieu of struts to support these structures instead.  A hint came from the study of vacuum Weyl solutions in \cite{Elvang:2002br} where static solutions composed of a vacuum bubble sandwiched between two black holes in $\mathbb{M}^4 \times$S$^1$  were constructed.  The two black holes are kept from colliding onto each other by the vacuum bubble instability to expand.

Building from \cite{Bah:2020pdz}, a new mechanism to construct and support charged solitons in the $\mathbb{M}^4 \times $T$^2$ is proposed in \cite{ourpaper}.  It is the most complete and non-trivial framework for building regular solitonic objects with the Weyl formalism. One circle in the torus is identified with the one in the five-dimensional setup of \cite{Bah:2020pdz} and the other is new.  Instead of having struts to support bubbles, they are replaced by rod sources where the new circle collapses as in \fig{multibubble} forming a chain of two species of bubbles.  The instability to expand for one species of rods provides the gravitational counterbalance between the neighboring rods of the other species \cite{Bah:2020pdz}.  This alternative pattern stabilizes the internal rods.  The overall structure and the edge rods are kept from expanding by adding electromagnetic fluxes. This is an interesting and rather counter-intuitive aspect of our mechanism. Indeed, in the Gibbons-Warner paradigm \cite{Gibbons:2013tqa} and in generic BPS microstate geometries, the role of electromagnetic fluxes is to avoid from collapsing and to counterbalance the gravitational attraction. Here, their role is strictly opposite by preventing from expanding and fluxes are not required to counterbalance the gravitational attraction.

The solutions obtained from the setup in \fig{multibubble} are smooth and geodesically complete. In this paper, we construct and study bubbling solutions that are made from a large number of smooth bubbles wrapped by fluxes, referred as \emph{Weyl Stars}. Their geometries are very curious as they exhibit a minimal S$^2$ that makes the spacetime to suddenly opens up before capping off to a region with non-trivial topology made by microscopic bubbles (see Fig.\ref{fig:WeylStarIntro}). We define such a feature as \emph{Bubble Bag End}.  

\begin{figure}[t]
\centering
\includegraphics[width=0.8\textwidth]{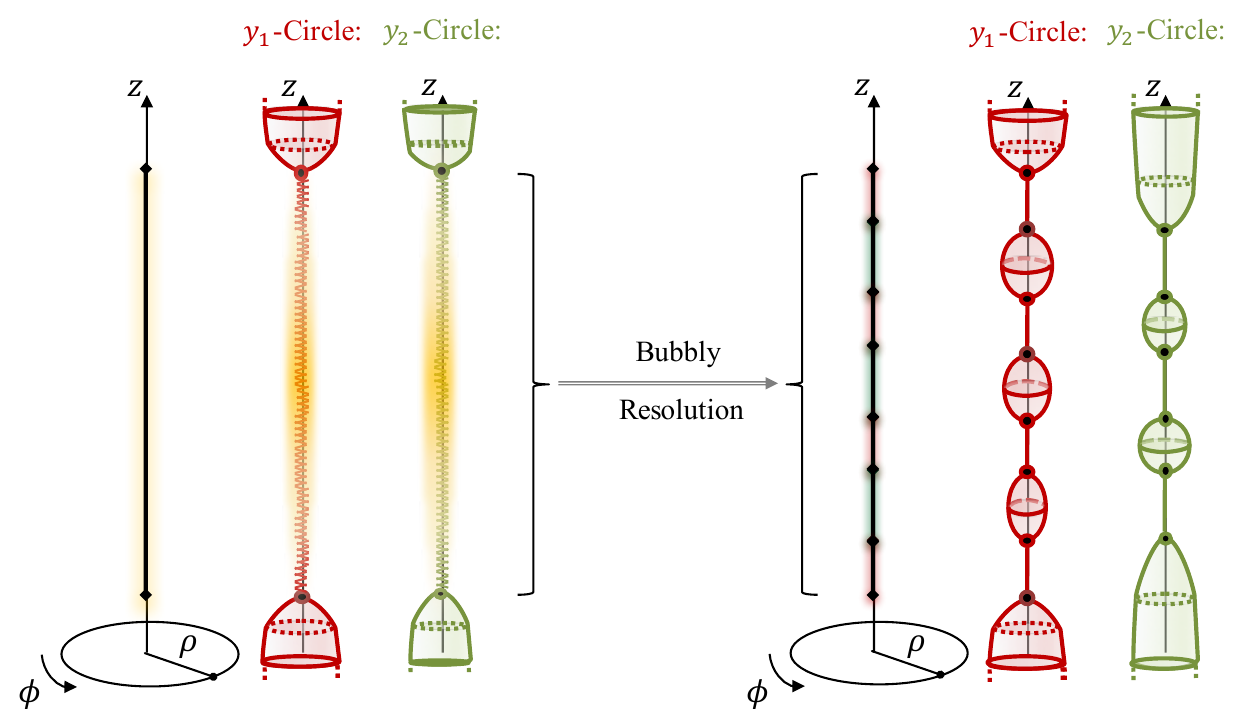}
\caption{In the first figure we consider a coarse graining of the rod sources to obtain a singular solution, dubbed bag spacetime.  The singularity is resolved by Weyl star geometries.}
\label{fig:BagResolution}
\end{figure}

So far we have described the framework for constructing smooth and regular solutions with the Weyl formalism.  It is also interesting to consider solutions where the full torus collapse instead of the alternating pattern of the circles as depicted in the first figure of \fig{fig:BagResolution}.  Not surprisingly, the solutions with these sources in $\mathbb{M}^4 \times $T$^2$ have a curvature singularity where the size of the surrounding S$^2$ diverges.  However, the smooth bag end spacetimes coincide with this singular solution in regions arbitrarily close to the singularity and resolve it entirely by capping off the spacetime at their bubbly surface. For these reasons we refer to the singular solutions as {\it Bag} spacetimes. \emph{We then have the first resolution of a curvature singularity with topological cycles wrapped by fluxes beyond the framework of supersymmetry and supergravity.}

\subsection*{Summary of results}

In this section, we summarize the results of the paper and provide a road map for the different sections.  We start in section \ref{sec:Weyl6d} with the explicit construction of the solitons and describe their smoothness and regularity conditions.  We also discuss various interesting limits of our solutions.  

To implement the mechanism outlined above, we consider Einstein-Maxwell theory with a two-form gauge field in six dimensions.  We have shown in \cite{ourpaper} that the Weyl formalism can be generalized to this context where the non-linear Einstein equations, with an appropriate Weyl ansatz, can be reduces to a set of linear equations that allows us to superpose different species of sources (see appendix \ref{App:Weyl6d} for a review).  This system can be embedded in IIB supergravity where the sources in our constructions correspond to non-BPS bound states of D1-D5-KKm systems.  

Our solutions describe \emph{ultra-compact axially-symmetric smooth horizonless solitons} in $\mathbb{M}^4 \times $T$^2$ that can live in the same parameter regimes as non-BPS and non-extremal black holes.  Upon KK reduction to four dimensions on the torus, the geometries correspond to asymptotically flat singular solutions of Einstein-Maxwell-Dilaton theories.  The four-dimensional singularities are resolved by uplifting to the six-dimensional or type IIB frame with the bubbling geometries.

The Weyl stars have a large number of rod sources where the algebraic constraints from regularity conditions are solvable.  The size of a soliton scales with the number of bubbles and with the asymptotic radii of the KK circles.  An important success here is that the stars can be made large without requiring conical defects at the bubbles, in contrast to the previous geometries of smooth non-BPS bubbling geometries constructed by the authors \cite{ourpaper,Bah:2020ogh,Bah:2020pdz}. 

As noted, the surface of Weyl stars is characterized by the degeneration pattern of the T$^2$ along the rod sources.  An important objective is to fully characterize this surface and the spacetime away from it.  This is the subject of section \ref{sec:WeylStar}.  We develop a procedure, using the Riemann sum approximation, that allows us to drastically simplify generic Weyl solutions a small distance away from the sources. With this method, we show that ``slightly above'' their surface Weyl star spacetimes are indistinguishable from bag spacetimes.

\begin{figure}[t]
\centering
\includegraphics[width=0.8\textwidth]{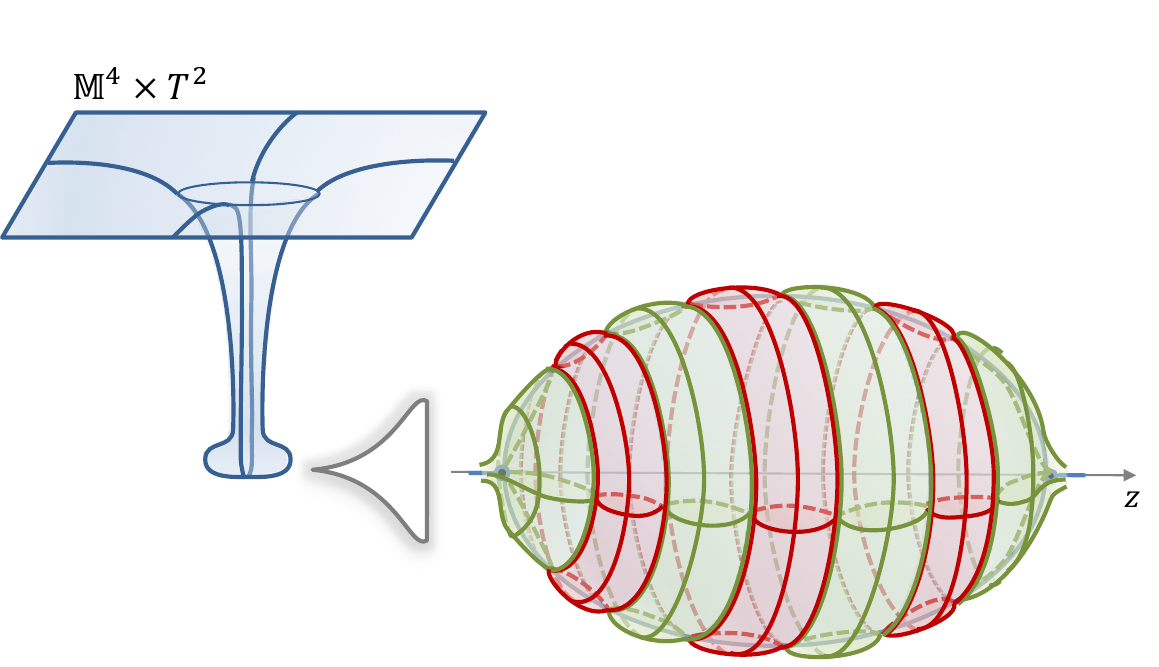}
\caption{Schematic description of a Weyl star spacetime. The T$^2$ circles are colored in red and green. The S$^2$ at the surface of the star is recovered by axisymmetric microscopic bubbles (7 in the figure) where one of the T$^2$ circles shrink smoothly.}
\label{fig:WeylStarIntro}
\end{figure}

To illustrate the bag in the summary, we consider a specific family of neutral Weyl stars, where the fluxes have been turned off.  The corresponding bag spacetimes that approximate the solutions above their surface are very simple solutions given by a single parameter $M$:
\begin{align}
ds_6^2 \=& -dt^2 \+ \sqrt{1-\frac{2M}{r}} \,\left( dy_1^2 +dy_2^2 \right) \+ \left(\frac{(r-M)^2-M^2 \cos^2\theta}{r(r-2M)} \right)^\frac{1}{4} \,\left(\frac{r \,dr^2}{r-2M} + r^2 \,d\theta^2 \right) \nonumber \\
&\+ r^2 \sin^2\theta \,d\phi^2\,, \nonumber
\end{align}
where $(y_1,y_2)$ parametrize the two additional compact dimensions. These bag spacetimes have not been studied before because they exhibit a naked singularity at $r=2M$. They have a T$^2$ that degenerates at the singularity, and more radically the size of the S$^2$, that surrounds the singularity in $\mathbb{M}^4$, diverges. As one falls toward the structure, the size of the S$^2$ is decreasing up to certain radius, $r\approx 2.1$M, where there is a minimal surface.  The space will then suddenly open up to a larger space similar to a ``bag'', and ultimately end at the singularity. A Weyl star has the same property above its surface: it has a minimal S$^2$ surface and suddenly opens to a larger space (see Fig.\ref{fig:WeylStarIntro}). However, the spacetime caps off smoothly at its surface by non-trivial, smooth and horizonless topological structure, like a \emph{bubble bag end}. 

Furthermore, we demonstrate that in the large mass limit, the Weyl star spacetime becomes more and more indistinguishable from the bag spacetime. We have the first examples, without the help of supersymmetry, of scaling smooth solutions that resemble a singular geometry up to an arbitrarily small distance above the singularity and resolve it into non-trivial smooth topologies.

Finally, we compare the basic properties of Weyl stars and bag spacetimes to four-dimensional black holes with the same conserved charges.  We show that the radius of the minimal S$^2$ of the former is between 1.5 and 2.3 times larger than that of the S$^2$ horizon, depending on the range of masses and charges.  Therefore, these solutions should be considered more as examples of ultra-compact objects build with ``topological matter'' than generic classical microstate geometries of non-extremal black holes.  However, since they use the same ingredients as the well-known BPS microstate geometries, i.e. non-trivial topologies and fluxes, we believe that turning on more degrees of freedom would allow them to scale more and more towards the black hole geometry, as classical microstate geometries of non-extremal black hole should do, if they exist.

In section \ref{sec:bagspacetimes}, we classify precisely the family of bag spacetimes that can be resolved by our smooth bubbling horizonless geometries. The neutral ones look generically like the metric above with one extra parameter, $D$, that we can identify with the relative contribution of one species of rods to the other in the corresponding Weyl star.  This parameter is also intimately related to the ratio of the asymptotic radii of the torus circles in the resolved spacetime.

\section{Bubbling Weyl solutions in six dimensions}
\label{sec:Weyl6d}

In this section, we review the work of \cite{ourpaper} where six-dimensional static axisymmetric solutions with two compact dimensions have been constructed using the Weyl formalism. We review the general framework in appendix \ref{App:Weyl6d}, and focus here on the construction of smooth bubbling solutions.  Our general setup is obtained from the following six-dimensional Einstein theory coupled to a two-form gauge field
\begin{equation}
 \,S_{6} \= \frac{1}{16 \pi G_6} \int d^6x \sqrt{-\det g}\,\left(R \-  \frac{1}{2} \,\left|F_3\right|^2\,\right)\,,
\label{eq:Action6d}
\end{equation}
where $F_3=dC_2$\footnote{ The norm is given as $
\left|F_3\right|^2 =  \frac{1}{3!}\,F_{3\,\mu\nu\sigma}\,F_3^{\,\mu\nu\sigma}\,.$} is the three-form field strength of the two-form gauge field in the theory.  

We restrict to solutions that are asymptotic to T$^2 \times\mathbb{M}^4$. The T$^2$ will be parametrized by $(y_1,y_2)$ with $2\pi R_{y_a}$ periodicities. The equations of motion are
\begin{equation}
d\star F_3 \= 0 \,,\qquad R_{\mu \nu} \= \frac{1}{2} \left(T_{\mu \nu} \- \frac{g_{\mu\nu}}{4} \,T_{\alpha}^{\,\,\alpha} \right)\,,
\label{eq:EinsteinMaxEq}
\end{equation}
where $T_{\mu\nu} = \frac{1}{2} \left[F_{3\,\mu\alpha\beta} F_{3\,\nu}^{\,\,\,\,\alpha\beta}- g_{\mu \nu}\,\left|F_3\right|^2\right]$ is the stress energy tensor of the gauge field.

\subsection{Weyl ansatz}
\label{sec:Weyl}

The ansatz that allows to build closed-form axisymmetric solutions is given by (see appendix \ref{App:Weyl6d} for a review)
\begin{align}
ds_{6}^2 = &\frac{1}{Z_1} \left[- W_0\,dt^2 + \frac{ dy_1^2}{W_0} \right] + \frac{Z_1}{Z_0}\, \left(dy_2 +H_0 \,d\phi\right)^2 + Z_0 Z_1\,\left[e^{2\nu} \left(d\rho^2 + dz^2 \right) +\rho^2 d\phi^2\right] ,\nonumber\\
 F_3 \= & d\left[ H_1 \,d\phi \wedge dy_2 \+ T_1 \,dt \wedge dy_1 \right]\,,\label{eq:WeylSol2circle}
\end{align}
where $(\rho,z,\phi)$ are the cylindrical Weyl's coordinates of an asymptotically-$\IR^3$ base and the warp factors $(Z_0,Z_1,W_0,\nu)$ and the gauge potentials $(H_0,H_1,T_1)$ are functions of $\rho$ and $z$. Therefore, the solutions can carry electric and magnetic line charges along the $y_1$-circle (in $T_1$ and $H_1$ respectively), and Kaluza-Klein magnetic charges (in $H_0$) along the $y_2$-circle.  \\

The solutions are given according to \emph{three harmonic functions} that solve a cylindrical Laplace equation for a flat three-dimensional base\footnote{The Laplacian operator $\cL$ can be also written in form formalism as $\star_3 d \star_3 d$ when it is applied on a scalar. Because all functions are not depending on $\phi$, $\nu$ does not enter in the equation and we have a Laplace equation in an apparent flat base.}
\begin{equation}
\cL \left(\log W_0 \right) \=\cL \left(L_0 \right)\=  \cL \left(L_1 \right) \= 0\,,\qquad \cL \equi \frac{1}{\rho} \,\partial_\rho \left( \rho \,\partial_\rho \right) + \partial_z^2 \,,
\label{eq:EOM1}
\end{equation}
The scalars $(Z_0, H_0)$ and $(Z_1, H_1,T_1)$ are given by\footnote{We have assumed self-duality $F_3$ in this paper for brevity.  In \cite{ourpaper}, one adds a constant parameter $q$ to move away from self-duality that corresponds to the ratio between electric and magnetic charges in $F_3$:
\begin{equation}
Z_1 = \cG^{(1)}_\ell\left(L_1\right) \,,\qquad \star_3 d\left(H_1\,d\phi\right) = \sqrt{\frac{2}{1+q^2}}\,dL_1\,, \qquad dT_1 =  q\sqrt{ \frac{2}{1+q^2}}\,\frac{dL_1}{ \cG^{(1)}_\ell(L_1)^2} \,,
\end{equation}}
\begin{equation}
Z_I \= \cG^{(I)}_\ell\left(L_I\right) \,,\qquad \star_3 d\left(H_I\,d\phi\right) \= dL_I\,, \qquad dT_1 \=   \frac{dL_1}{ \cG^{(1)}_\ell(L_1)^2} \,,
\label{eq:WGF&H}
\end{equation}
where $\star_3$ is the Hodge star operator with respect to the three-dimensional base, and $ \cG^{(I)}_\ell$ is chosen from one of the two generating functions of one variable given by\footnote{There are four possible generating functions  \cite{ourpaper}, but only two are considered in this article.}
\begin{equation}
 \cG^{(I)}_1 (x) \=\frac{\sinh(a_I \,x+b_I)}{a_I}\,,\qquad  \cG^{(I)}_2(x)\= x +b_I \,,\qquad a_I \in \mathbb{R}_+, \,b_I\in \IR \,.
\label{eq:DefFi}
\end{equation}
Note that $ \cG^{(I)}_2$ can be obtained from $ \cG^{(I)}_1$ by taking $a_I \to 0$ and rescaling $b_I$ appropriately.

\noindent The function $\nu$ satisfies the first order equations: 
\begin{align}
& \frac{1}{\rho}\partial_z \nu = \frac{1}{2}\, \partial_\rho \log W_0\,\partial_z \log W_0+ \frac{a_0^2}{2}\,\partial_\rho L_0\,\partial_z L_0 + a_1^2\,\partial_\rho L_1\,\partial_z L_1, \label{eq:nuEq} \\
& \frac{1}{\rho}\partial_\rho \nu = \frac{1}{4} \left( \left(\partial_\rho \log W_0\right)^2- \left(\partial_z \log W_0 \right)^2\right)+ \frac{a_0^2}{4} \left( \left( \partial_\rho L_0 \right)^2- \left(\partial_z L_0 \right)^2\right)  + \frac{a_1^2}{2} \left( \left( \partial_\rho L_1\right)^2- \left(\partial_z L_1 \right)^2\right),\nonumber
\end{align}
where $a_I=0$ if $ \cG^{(I)}_\ell =  \cG^{(I)}_2$.  The integrability of these equations is guaranteed by the harmonic equations in \eqref{eq:EOM1}.  

It is important to note that $W_0$ is a pure ``vacuum'' warp factor, as it is not related to any gauge potentials, whereas the $Z_I$ are generated and induced by the gauge fields. However, one can take a neutral limit by sending all gauge potentials to zero while keeping $Z_I$ non-trivial, which then satisfies the vacuum equation $\cL (\log Z_I) =0$. The neutral limit is possible only for the case $\ell=1$ in \eqref{eq:WGF&H} by sending $b_I \to \infty$ with $e^{b_I}/a_I$ and $a_I L_I$ held fixed.\\

We have defined a family of axisymmetric static Weyl solutions with a very large phase space. In \cite{ourpaper} and in the appendix \ref{sec:BPSlimitgen}, we have shown that taking $\ell=2$ and sourcing the harmonic functions by point particles corresponds to multicenter BPS solutions.

As discussed in the appendix \ref{sec:nonBPSgen}, generic Weyl solutions are obtained by taking $\ell=1$ and sourcing the harmonic functions by \emph{rods} on the $z$-axis. The physical rod sources consist of six-dimensional embedding of four-dimensional \emph{non-extremal black holes}, and \emph{two species of bubbles, species-1 and species-2, where either the $y_1$ or $y_2$ circle degenerates smoothly}. Furthermore, if the rods are not connected, strings with negative tension, or \emph{struts}, appear to prevent the structure from collapsing. These struts disappear when the rod sources touch, and the structure is prevented from collapsing by pure topology \cite{ourpaper}. Therefore, unlike the BPS regime,  electromagnetic fluxes are not necessary to construct Weyl solutions and one can even take a neutral limit without changing the topology \cite{ourpaper}.  The flux, however, can play another crucial role.  Vacuum KK bubbles mediate a decay of the spacetime where they expand to eat up all the gravitational background \cite{Witten:1981gj}.  The presence of the electromagnetic flux provides different asymptotic boundary conditions of spacetime with conserved charges and therefore can stabilize the overall bubble structure from expanding \cite{Stotyn:2011tv,ourpaper}.

In this paper, we are interested in smooth non-BPS bubbling geometries consisting of a succession of species-1 and species-2 bubbles on a line with an odd number of bubbles. We will review this specific class of solutions in the next subsection.

\subsection{Smooth bubbling Weyl solutions}
\label{sec:smoothbubble}

\begin{figure}[t]
\centering
\includegraphics[width=0.8\textwidth]{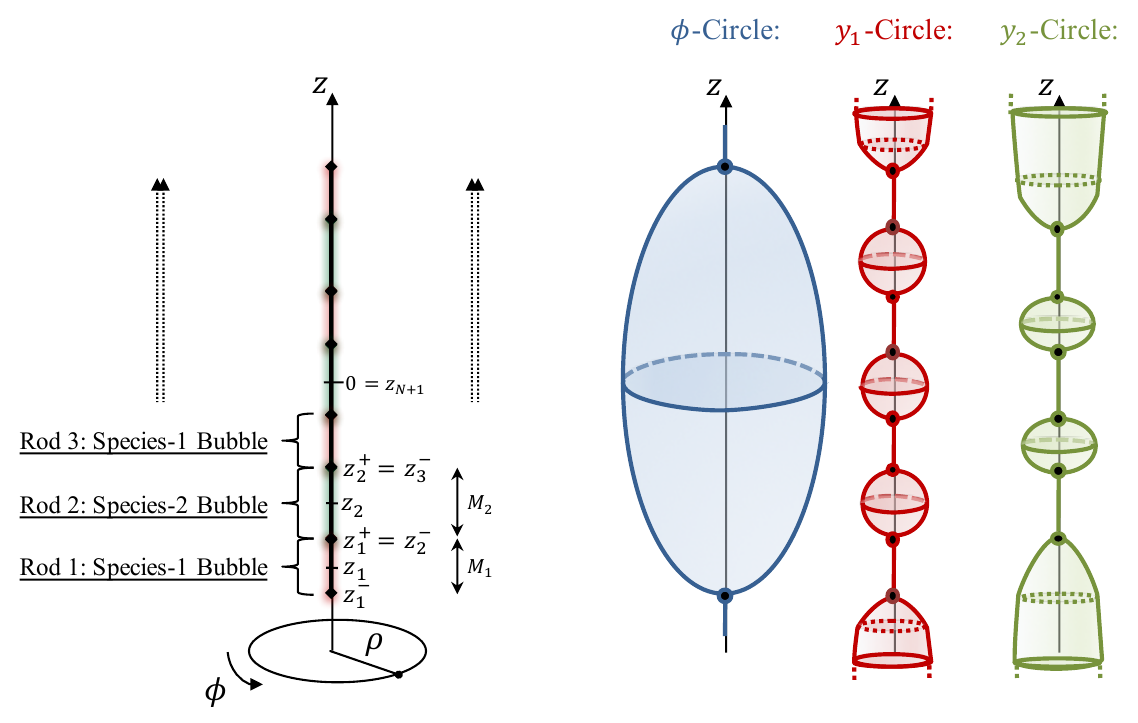}
\caption{Rod profile of the $n$-bubble solutions and the behavior of the circles on the $z$-axis.}
\label{fig:nTouchingRods}
\end{figure}

We consider the branch of solutions with $\ell=1$ \eqref{eq:WGF&H}. The harmonic functions are sourced by $n=2N+1$ touching rods of length $M_i>0$ and centered on $z=z_i$. Without loss of generality, we can order them as $z_i < z_j$ for $i<j$ and fix the origin of the $z$-axis such that $z_{N+1}=0$. Our conventions are illustrated on the left side of Fig.\ref{fig:nTouchingRods}. The coordinates of the rod endpoints on the $z$-axis are given by
\begin{equation}
z^\pm_i \equi z_i \pm \frac{M_i}{2}\= \sum_{j=1}^{i} M_j \- \sum_{j=1}^{N+1} M_j + \frac{M_{N+1}-(1\mp 1)M_i}{2}\,.
\label{eq:zintouchingVac}
\end{equation}
We define the distances to the endpoints $r_\pm^{(i)}$ and the generating functions $E_{\pm \pm}^{(i,j)}$ such as
\begin{equation}
r_\pm^{(i)} \equiv \sqrt{\rho^2 + \left(z-z^\pm_i\right)^2}\,, \qquad E_{\pm \pm}^{(i,j)} \equiv r_\pm^{(i)} r_\pm^{(j)} + \left(z-z_i^{\pm}  \right)\left(z-z_j^{\pm}  \right) +\rho^2,
\label{eq:Rpmdef}
\end{equation}
The harmonic functions sourced at the rods are given by 
\begin{equation}
\log W_0 =  \sum_{i}^n G_i\,\log \left(  \frac{r_+^{(i)}+r_-^{(i)}+ M_i}{r_+^{(i)}+r_-^{(i)}- M_i}\right)\,, \quad L_I = \sum_{i}^n P_i^{(I)}\,\log \left(   \frac{r_+^{(i)}+r_-^{(i)}+ M_i}{r_+^{(i)}+r_-^{(i)}- M_i}\right),
\label{eq:harmonicfuncGen}
\end{equation}
where $(G_i,P_i^{(0)},P_i^{(1)})$ are specific weights associated to each rod that depend whether the $i^\text{th}$ rod corresponds to a black hole \eqref{eq:BlackBraneGicharged} or a species-1 bubble \eqref{eq:1TSGi} or a species-2 bubble \eqref{eq:2TSGi}. We therefore consider the weights that correspond to a configuration of species-1 and species-2 bubbles as depicted in Fig.\ref{fig:nTouchingRods}:
\begin{equation}
\sinh b_1\, P_{2i-1}^{(1)} = \sinh b_0\, P_{2i-1}^{(0)} = G_{2i-1} = \frac{1}{2}\,,\qquad \sinh b_0\, P_{2i}^{(0)} =  1 \,,\quad P_{2i}^{(1)} = G_{2i} =0\,.
\end{equation}
We can now insert the expressions of the harmonic functions to derive the warp factors and gauge potentials \eqref{eq:WGF&H} and \eqref{eq:nuEq} with the condition $a_I = \sinh b_I$ necessary to have an $\mathbb{M}^4 \times $T$^2$ asymptotic spacetime (see appendix \ref{sec:nonBPSgen} for more details). Note that $a_I\geq 0$ \eqref{eq:DefFi} restricts the gauge-field parameters $b_I$ to be
\begin{equation}
b_I \geq 0\,.
\label{eq:bIrestriction}
\end{equation} 
Finally, smooth bubbling Weyl solutions can be recast such that
\begin{align}
&ds_6^2  = \frac{1}{\mathcal{Z}_1} \left[- dt^2 + U_1 dy_1^2 \right] +  \frac{U_2\,\mathcal{Z}_1}{\mathcal{Z}_0}\,(dy_2+H_0\,d\phi)^2 + \frac{\mathcal{Z}_0\mathcal{Z}_1}{U_1 U_2}\left[e^{2\nu}\,\left(d\rho^2+dz^2 \right) +\rho^2 d\phi^2\right], \nonumber\\
&F_3 =dH_1 \wedge d\phi\wedge dy_2 \+  dT_1  \,\wedge\, dt \,\wedge\, dy_1 \,,
\label{eq:met&GFbubbling}
\end{align}
where the quantities,\footnote{The functions here are related to the ones in \eqref{eq:WeylSol2circle} as $$\mathcal{Z}_1 = \frac{Z_1}{W_0}, \qquad U_1 = \frac{1}{W_0^2}, \qquad \frac{\mathcal{Z}_0}{U_2} = \frac{Z_0}{W_0} $$} which are independent of the gauge fields, are given by
\begin{align}
U_1 \=& \prod_{i=1}^{N+1} \left( 1- \frac{2M_{2i-1}}{r^{(2i-1)}_+ +r^{(2i-1)}_-+M_{2i-1}} \right)\,,\qquad U_2 \= \prod_{i=1}^N \left( 1- \frac{2M_{2i}}{r^{(2i)}_+ +r^{(2i)}_-+M_{2i}}\right)\,, \nonumber \\
e^{2\nu} \=& \frac{E_{-+}^{(1,n)}}{\sqrt{E_{++}^{(n,n)}E_{--}^{(1,1)}}} \,\prod_{i=1}^{N}\prod_{j=1}^{N+1} \sqrt{\frac{E_{--}^{(2i,2j-1)}E_{++}^{(2i,2j-1)}}{E_{-+}^{(2i,2j-1)}E_{+-}^{(2i,2j-1)}}}\,,\label{eq:WarpFactorsVacMultiple}
\end{align}
and the ones depending on the gauge fields are
\begin{align}
\mathcal{Z}_0&\= \frac{e^{b_0} -e^{-b_0}\, U_1 \,{U_2}^2}{2 \,\sinh b_0}\,,\qquad \mathcal{Z}_1 \= \frac{e^{b_1} -e^{-b_1}\, U_1 }{2 \,\sinh b_1}\,, \nonumber\\
H_0 & \=  \frac{1}{2\sinh b_0} \,\left[r_-^{(1)}-r_+^{(2N+1)} \+ \sum_{i=1}^N (r_-^{(2i)}-r_+^{(2i)}) \right]\,,\label{eq:WarpFactorsVacMultipleCharged}\\
H_1 & \= \frac{1}{2\sinh b_1}\,  \sum_{i=1}^{N+1} (r_-^{(2i-1)}-r_+^{(2i-1)})\,,\quad T_1 \=  -\sinh b_1 \,\, \frac{e^{b_1} +e^{-b_1} \, U_1 }{e^{b_1} -e^{-b_1} \, U_1 }\,,\nonumber
 \end{align}

It follows from the expression of the rod-endpoint distances \eqref{eq:Rpmdef} that $0\leq U_I <1$ and that $U_1$ and $U_2$ vanish at the odd and even rods respectively. Therefore, the $y_1$ and $y_2$ circles indeed alternatively shrink on the segment of the $z$-axis along the sources as depicted on the right-hand side of Fig.\ref{fig:nTouchingRods}. 
The regularity of these degeneracies will impose $n=2N+1$ \emph{bubble equations} that will constrain the local geometry with respect to the extra-dimensional radii $R_{y_a}$. These equations will be made clearer for our bubbling configuration in the next section.

\subsection{Regularity and bubble equations}

The solutions are regular out of the $z$-axis and asymptote to T$^2\times\mathbb{M}^4$. Moreover, the $z$-axis is an axis of rotational symmetry where at least one of the three circles shrink for any $z$. Therefore, the spacetime is geodesically complete for $\rho \geq 0$ and $z\in \mathbb{R}$. More precisely, the local geometry on the $z$-axis depends on $z$ such as:
\begin{itemize}
\item[•] For the semi-infinite segments above and below the bubble configuration, $z\geq z_{2N+1}^+$ and $z\leq z_{1}^-$, $U_1,\,U_2$ and $\mathcal{Z}_I$ are finite while $e^{2\nu}=1$. Therefore, the $\phi$-circle shrinks smoothly as the origin in cylindrical coordinate system.  One can also check that the three-form flux, $F_3$, is regular in this region.  
\item[•] The bubbles are located on the segment $z_1^- \leq z \leq z_{2N+1}^+$. At each individual segment $z_i^-  \leq z \leq z_{i}^+$, we have $U_1 \propto \rho^2$ and $U_2$ finite if $i$ is odd or $U_2 \propto \rho^2$ and $U_1$ finite if $i$ is even, and $e^{2\nu}\propto \rho^2$ in both cases. Therefore, the $\phi$-circle have a finite size in these regions (see Fig.\ref{fig:nTouchingRods}). On the other hand, the $y_1$-circle ($y_2$-circle) shrink if $U_1$ ($U_2$) vanish.  We require each rod region to be regular and allow for conical defects associated to orbifold parameters $k_i\in \mathbb{N}$.  This leads to $2N+1$ conditions\footnote{We have used a notation such that $\prod_i^j (\ldots)=1$ if $i>j$.} referred to as \emph{bubble equations} and given as (see \eqref{eq:RegRy1} and \eqref{eq:RegRy2} for general bubble systems)  
\begin{align}
& \frac{2}{k_1}\,M_1 \prod_{j=1}^{N} \sqrt{1+\frac{M_{2j}}{\sum_{\ell=1}^{2j-1}M_\ell}}\,\prod_{j=1}^{N} 1+\frac{M_{2j+1}}{\sum_{\ell=1}^{2j}M_\ell} \= \widetilde{R}_{y_1}\,,\nonumber\\
& \frac{2}{k_i} \,\sqrt{M_i(M_i+M_{i-1})}\,\prod_{j=1}^{i-1}\prod_{k=i+1}^{2N+1} \,\left( \left( 1+\frac{M_{j}}{\sum_{\ell=j+1}^{k}M_\ell}\right)\left(1-\frac{M_{j}}{\sum_{\ell=j}^{k-1}M_\ell} \right)\right)^{\alpha_{jk}} \label{eq:RegRyconnected}\\
&\hspace{1cm} \times \prod_{j=1}^{i-1} \left(1+\frac{M_{j}}{\sum_{\ell=j+1}^{i}M_\ell} \right)^{\alpha_{ij}} \prod_{j=i+1}^{2N+1} \left(1+\frac{M_{j}}{\sum_{\ell=i}^{j-1}M_\ell} \right)^{\alpha_{ij}}  \= \begin{cases}
\widetilde{R}_{y_1} \quad \text{if }i>1\text{ is odd,}   \\
\widetilde{R}_{y_2} \quad \text{if }i>1\text{ is even,}
\end{cases}\nonumber
\end{align}
where $\alpha_{jk}$ is $1$ if $j$ and $k$ have the same parity or $1/2$ otherwise, and $(\widetilde{R}_{y_1},\widetilde{R}_{y_2})$ are the extra-dimensional radii rescaled by the gauge field parameters:
\begin{equation}
\widetilde{R}_{y_1} \equi \frac{2 \sinh b_1}{e^{b_1}} \sqrt{\frac{2 \sinh b_0}{e^{b_0}} }\,R_{y_1}\,,\qquad \widetilde{R}_{y_2} \equi \frac{2 \sinh b_0}{e^{b_0}}\,R_{y_2}\,.
\label{eq:tildeRy}
\end{equation}
There are $2N+1$ bubble equations for $2N+1$ rod lengths, $M_i$; all rod lengths are then fixed according to the orbifold parameters and the rescaled radii $\widetilde{R}_{y_a}$. The unconstrained internal degrees of freedom are then the orbifold parameters, the gauge field parameters $b_I$ and the number of bubbles.

We refer the reader to \cite{ourpaper} for a complete analysis of the local topology of the solutions at the bubble loci. In short, all the bubbles are sitting at the origin of a $\IR^2/\mathbb{Z}_{k_i}$, and the two bubbles at the ends of the configuration have an S$^3$ topology while those in the middle have an S$^2\times$S$^1$ topology.
\end{itemize}
We are free to fix the values of the orbifold parameters, $k_i$, as long as they are quantized. We will refer to individual bubbles that are free from conical defects, $k_i=1$, as \emph{microscopic bubbles}. The term ``microscopic'' is apt since the bubble equations constrain $M_i \leq R_{y_a}$ if $k_i =1$ \cite{ourpaper}.  When $k_i>1$, it rescales $\widetilde{R}_{y_a}$ in the bubble equations and therefore allows for larger solutions of $M_i$ \cite{ourpaper}. 

\noindent In principle, smooth solutions exist if and only if the bubbles are microscopic. However, geometries with conical defects are usually considered as smooth in string theory. On one hand, conical defects have a well-known classical resolution by blowing up smooth Gibbons-Hawking cycles \cite{GIBBONS1978430,Bena:2007kg}. On the other hand, in AdS/CFT the corresponding worldsheet conformal field theory is completely well-defined and there are many examples, for instance in AdS$_3$/CFT$_2$ \cite{Jejjala:2005yu,Giusto:2004id,Giusto:2012yz}, where regular CFT states have bulk duals with conical defects.

In \cite{ourpaper}, three-bubble solutions with potential conical defects at the bubbles have been constructed. This example can be derived from \eqref{eq:met&GFbubbling} and was appropriate to illustrate the physics of the replacement of struts by smooth bubbles in six dimensions. Moreover, the bubble equations were completely solvable with $M_i = \cO( k_i \,\widetilde{R}_{y_a})$. Thus, one needs large conical defects, $k_i \gg 1$, to have a rod configuration larger than the KK scales. 

We aim to construct solutions with a large number of bubbles and investigate their physics. In particular, we can ask whether increasing the number of bubbles can avoid taking large orbifold parameters, or even allow solutions with large structure made from microscopic bubbles. We will refer to them as \emph{Weyl star}. The main difficulty will be to solve the bubble equations generically and to simplify the form of the warp factors \eqref{eq:WarpFactorsVacMultiple} and \eqref{eq:WarpFactorsVacMultipleCharged} at large $N$. 

 First, we describe the solutions from a four-dimensional point of view after compactification on the T$^2$, some interesting limits of the gauge field parameters $b_I$ and the type IIB embedding of the solutions as D1-D5-KKm bubbling geometries in the non-BPS regime.

\subsection{Kaluza-Klein reduction and conserved charges}
\label{sec:KKreduction}

In this section, we discuss the compactifications to four dimensions, and  consider a truncation that keeps only the non-trivial fields of the solutions.  

\subsubsection*{Reduction to four dimensions}

The compactifications on $y_2$ and $y_1$ give rise to a scalar and KK gauge field $(\Phi_2,H_0d\phi)$ and a scalar $\Phi_1$ respectively. We obtain a four-dimensional Einstein-Maxwell-dilaton action 
\begin{align}
S_{4}^\text{KK} &=  \frac{1}{\left(16 \pi G_4\right) } \int d^4x \sqrt{-\det g}\,\left( R - \mathcal{L}_4 \right), \\
\mathcal{L}_4 &= \frac{1}{2}\,\partial_\mu \Phi_2 \,\partial^\mu \Phi_2 + \,\frac{1}{2}\,\partial_\mu \Phi_1 \,\partial^\mu \Phi_1 - \frac{e^{\frac{-2\sqrt{2}}{\sqrt{3}}\,\Phi_2-\frac{1}{\sqrt{3}}\,\Phi_1}}{2}\,\left| F^{(m0)}\right|^2  \nonumber \\
&+ \frac{e^{\frac{\sqrt{2}}{\sqrt{3}}\,\Phi_2-\frac{1}{\sqrt{3}}\,\Phi_1}}{2}\,\left| F^{(m1)}\right|^2 + \frac{e^{-\frac{\sqrt{2}}{\sqrt{3}}\,\Phi_2+\frac{1}{\sqrt{3}}\,\Phi_1}}{2}\,\left| F^{(e1)}\right|^2. \nonumber 
\end{align}
with $G_4 \equiv \frac{G_6}{(2\pi)^2 R_{y_1}R_{y_2}} $. Note that we have turned off a KK vector associated to $y_1$ and the components of $F_3$ orthogonal to $y_1$ and $y_2$ in the truncation. The smooth bubbling geometries \eqref{eq:met&GFbubbling} are solutions of this action with
\begin{equation}
\begin{split}
ds_4^2 &\=-\sqrt{\frac{U_2 U_1}{\mathcal{Z}_0 \mathcal{Z}_1^2}} \,dt^2  \+\sqrt{\frac{\mathcal{Z}_0 \mathcal{Z}_1^2}{U_2 U_1}} \,\left[e^{2\nu} \left(d\rho^2 + dz^2 \right) +\rho^2 d\phi^2\right]\,,\\
e^{\frac{1}{\sqrt{3}}\,\Phi_1} &\= \frac{1}{\sqrt{U_1}}\,\left(\frac{\mathcal{Z}_1^2 \mathcal{Z}_0}{U_2}\right)^\frac{1}{6}\,,\qquad e^{\frac{1}{\sqrt{6}}\Phi_2}  \= \left(\frac{\mathcal{Z}_0}{U_2\,\mathcal{Z}_1}\right)^{\frac{1}{3}}\,, \\
 F^{(m0)}& \= dH_0 \wedge d\phi\,,\qquad F^{(m1)} \= dH_1 \wedge d\phi\,,\qquad  F^{(e1)} \= dT_1 \wedge dt  \,.
\label{eq:4dFrameworkcharged}
\end{split}
\end{equation} The four-dimensional metric and the scalars are singular in the region of the rods as it can be checked with \eqref{eq:WarpFactorsVacMultiple} and \eqref{eq:WarpFactorsVacMultipleCharged}.  Indeed, these singularities are genuine curvature singularities that signal the breakdown of the four-dimensional system.  These are resolved by adding the degrees of freedom that uplift the system to the six-dimensional framework.

\subsubsection*{Conserved charges}

To compute the conserved charges, we expand the solution at large distance $r\to \infty$, where $r$ is the asymptotic spherical coordinates, $ 
\rho \equi r \, \sin \theta \,,\,\, z \equi r \, \cos \theta$. With the conventions of \cite{Myers:1986un}, the charges can be read from the expansion as follows
\begin{align}
\sqrt{\frac{U_2 U_1}{\mathcal{Z}_0 \mathcal{Z}_1^2}} &\,\sim\, 1 \- \frac{2G_4 \cM}{r}\,,\qquad F^{(mI)} \sim -\sqrt{16\pi G_4}\,\cQ_m^{(I)} \,\sin\theta  \,d\theta \wedge d\phi \+ \ldots\,,\nonumber\\
 F^{(e1)} &\sim  \frac{\sqrt{16\pi G_4}\,\cQ_e^{(1)}}{r^2} \,dt\wedge dr \+ \ldots\,,
 \label{eq:AsymptoticExpGen}
\end{align}
where $\cM$ is the four-dimensional ADM mass, $\cQ_e^{(1)}$ is the electric charge, $\cQ_m^{(0)}$ and $\cQ_m^{(1)}$ are the magnetic charges. We find for our solutions
\begin{align}
\cM &= \frac{1}{4 G_4}\,\left[\coth b_0 \,\sum_{i=1}^{N} M_{2i}\+ \frac{\coth b_0 +2\coth b_1 -1}{2} \sum_{i=1}^{N+1} M_{2i-1} \right]\,,\label{eq:conservedchargesrod}\\
 \cQ_e^{(1)}&=\cQ_m^{(1)} =  \frac{1}{8\sqrt{\pi G_4}\,\sinh b_1} \sum_{i=1}^{N+1} M_{2i-1}\,, \quad \cQ_m^{(0)}  =  \frac{1}{8\sqrt{\pi G_4}\,\sinh b_1} \left[ \sum_{i=1}^{2N+1} M_i \+ \sum_{i=1}^N M_{2i} \right]\,. \nonumber
\end{align}

Note that the electric and magnetic charges that descend from $F_3$, $\cQ_e^{(1)}$ and $\cQ_m^{(1)}$, are equal due to the self-duality of $F_3$. This assumption can be lifted and one can allow a general SO(2) rotation between both charges as discussed in \cite{ourpaper}.

We remind the reader that $b_I$ are restricted to positive or null values \eqref{eq:bIrestriction}. There are two interesting limits to consider: $b_I \to 0$ and $b_I \to \infty$.  

In the $b_I\to 0$ limit, the conserved charges seem to diverge. However, it can be verified from the bubble equations \eqref{eq:RegRyconnected} that all rod lengths, $M_i$, also go zero as $\sinh b_I$ from the $\tilde{R}_{y_a}$ dependence. Therefore, the conserved charges can remain finite. We will see in a next section that the complete structure shrinks to a point in this limit and smoothly transits to a BPS Taub-NUT space.

For $b_I \to \infty$, nothing special happens to the bubble equations and the lengths of the rods remain finite. However, the electric and magnetic charges clearly vanish. As argued at the end of the section \ref{sec:Weyl}, this limit corresponds to a smooth neutral limit which preserves the bubbling geometry of the solutions.

\subsection{Neutral and BPS limits}

\subsubsection*{A neutral limit: smooth chain of vacuum bubbles}
\label{sec:neutrallim}

We consider the limit $b_I \to \infty$. From \eqref{eq:WarpFactorsVacMultipleCharged}, the gauge potentials $(H_I,T_1)$ and the warp factors $\mathcal{Z}_I$ simplify to constants, and the solutions dramatically reduce to 
\begin{equation}
ds_6^2  \= - dt^2 \+ U_1\, dy_1^2 \+ U_2 \,dy_2^2 \+ \frac{1}{U_1 U_2}\left[e^{2\nu}\,\left(d\rho^2+dz^2 \right) +\rho^2 d\phi^2\right], \quad F_3 \= 0\,,
\label{eq:neutralWeylmetric}
\end{equation}
where $U_1$, $U_2$ and $\nu$ are the same as in \eqref{eq:WarpFactorsVacMultiple}. Because $U_1$ and $U_2$ are the main factors that determine the topology, the $y_1$ and $y_2$ circles alternatively shrink at the rods, and the bubbling geometries are preserved. Therefore, the neutral solution consists of a sequence of alternating species of vacuum KK bubbles.

The regularity constraints are still given by \eqref{eq:RegRyconnected} with $\widetilde{R}_{y_a} \to R_{y_a}$ \eqref{eq:tildeRy}. An important observation is that since $\widetilde{R}_{y_a} \leq R_{y_a}$, the rod lengths, $M_i$, and the size of the whole structure are bigger without fluxes. This is rather counter-intuitive as it is generally thought, from the BPS multicenter solutions, that electromagnetic fluxes are necessary to compensate for gravitational attraction, and that they then have rather a repulsive role. For Weyl solutions, electromagnetic fluxes seem to have the opposite effect by compressing the geometry. As argued in \cite{ourpaper}, electromagnetic fluxes have an attractive role in preventing bubbles from expanding and the solutions are always prevented from collapsing gravitationally, regardless of the fluxes. The interpretation is that KK bubbles are hardly compressible because they want to expand. This gives a pressure of its own that counterbalances the gravitational attraction.

\subsubsection*{A BPS limit: Taub-NUT space}
\label{sec:BPSlim}

We now take $b_I\to 0$. In order to find non-trivial solutions to the bubble equations \eqref{eq:RegRyconnected}, we also consider the specific limit $b_1 \= \sqrt{b_0} \to 1$.\footnote{Taking $b_1 \propto b_0^x$ with $x\geq  \frac{1}{2}$ gives a similar result \cite{ourpaper}.}  In this limit, $\widetilde{R}_{y_1} \sim \widetilde{R}_{y_2} \to 0$ and the rod lengths behaves as $M_i \= b_0 \,q_i$, where $q_i$ are finite and positive. Therefore, the whole rod structure smoothly shrinks to a point located at $\rho=0$ and $z=0$.

\noindent We now expand the solutions \eqref{eq:met&GFbubbling} at leading order in $b_0$, and we obtain
\begin{align}
ds_6^2  = &- dt^2 +  dy_1^2+  \frac{1}{1+\frac{Q_0}{\sqrt{\rho^2+z^2}}}\,\left(dy_2+ \frac{z\,Q_0}{\sqrt{\rho^2+z^2}}\,d\phi\right)^2 \\
&+\left( 1+\frac{Q_0}{\sqrt{\rho^2+z^2}} \right)\,\left[d\rho^2+dz^2+\rho^2 d\phi^2\right], \qquad F_3 =0 \,,
\label{eq:met&GFbubbling}
\end{align}
where 
\begin{equation}
Q_0 \equi \frac{1}{2}\sum_{i=1}^{2N+1} q_i, \qquad \mbox{with} \qquad q_i = \lim_{b_0 \to 0} \frac{M_i}{b_0}.  
\end{equation}
The solution is $\mathbb{R}_t \times S^1_{y_1} \times \text{TN}_{k}$, where $\text{TN}_{k}$ is a Taub-NUT space with charge $k =\frac{Q_0}{R_{y_2}}$. It is smooth everywhere and the time slices at $(\rho,z)=(0,0)$ are a S$^1$ fibration over $\mathbb{R}^4/\mathbb{Z}_{k}$.    

It is quite remarkable that the limit $b_I \to 0$ has changed the topology smoothly: it has transformed a chain of bubbles at some origins of an orbifold of $\IR^2$ into an origin of an orbifold of $\IR^4$. By reversing the arguments, the parameters $b_I$ have blown up a Taub-NUT space into non-trivial non-BPS bubbling geometries. They have moved away non-trivially in the phase space of Einstein solutions, from the BPS regime to the non-BPS one, while keeping the solutions smooth all along the flow.

\subsection{Type IIB embedding: D1-D5-KKm solutions}
\label{sec:TypeIIB}

The six-dimensional framework detailed in section \ref{sec:Weyl6d} can be seen as the minimal pure $\cN=(1,0)$ six-dimensional supergravity. This theory arises as a consistent truncation of type IIB string theory on T$^4$ \cite{Chow:2014cca}. 

Our Weyl solutions, given by the six-dimensional ansatz in section \ref{sec:Weyl}, can be embedded in type IIB by considering
\begin{equation}
ds_{\text{IIB}}^2 = ds_6^2 \+ ds(T^4)^2 \,,\quad C^{(2)} =  H_1 \,d\phi \wedge dy_2 + T_1 \,dt \wedge dy_1,
\end{equation}
where $C^{(2)}$ is the $2$-form R-R gauge fields.  All other IIB potentials and dilaton vanish. 

From a type IIB perspective, the solutions can carry D1-brane charges in $T_1$, D5-brane charges in $H_1$ and KKm charges in $H_0$. The D1 and D5 charges must be equal for $F_3=dC^{(2)}$ to be self-dual in six dimensions and the dilaton to be trivial.

With that respect, the branch of BPS solutions given by $\ell=2$ in \eqref{eq:WGF&H} corresponds to static extremal D1-D5-KKm black holes and static Gibbons-Hawking centers on a line (see \cite{ourpaper} or the appendix \ref{sec:BPSlimitgen}). The branch of non-BPS solutions given by $\ell=1$ corresponds to non-extremal static D1-D5-KKm black holes, smooth D1-D5-KKm bubbles and KKm bubbles on a line (see \cite{ourpaper} or the appendix \ref{sec:nonBPSgen}).

Therefore, the smooth bubbling Weyl solutions constructed in section \ref{sec:smoothbubble} corresponds to a chain of static non-BPS D1-D5-KKm bubbles and KKm bubbles. The conserved charges are given by \eqref{eq:conservedchargesrod} where $\cQ_e^{(1)}$ and $\cQ_m^{(1)}$ are the equal D1 and D5 brane charges, while $\cQ_m^{(0)}$ corresponds to the KKm charge. One can dial these charges with the gauge field parameters $b_I$, from the neutral limit of section \ref{sec:neutrallim} where the bubbles become vacuum KK bubbles to the BPS limit of section \ref{sec:BPSlim} where the bubbles shrink to a smooth KKm Taub-NUT center.

\section{Weyl star: a bubble bag end}
\label{sec:WeylStar}

In the previous section, we have seen how smooth six-dimensional bubbling solutions can be generically constructed using the Weyl formalism.  In this section, we construct smooth Weyl solutions with large number of connected bubbles, $N\gg1$, that we refer to as \emph{Weyl Stars}. We will mostly focus on solutions made of a large number of microscopic bubbles free from conical defects, $k_i=1$, i.e. on solutions that are entirely free from conical singularities.

The basic rod setup for the Weyl star is depicted in Fig.\ref{fig:nTouchingRods}, and the solutions are given by the metric and gauge field \eqref{eq:met&GFbubbling} with \eqref{eq:WarpFactorsVacMultiple} and \eqref{eq:WarpFactorsVacMultipleCharged}. The internal degrees of freedoms characterized by the $M_i$'s are fixed according to the radii of the extra dimensions by the bubble equations  \eqref{eq:RegRyconnected}. These equations are the key conditions that constrain the physics of the bubbling solutions.  For the ease of simplicity, we will focus on solutions with $\widetilde{R}_{y_1} \= \widetilde{R}_{y_2} \= \widetilde{R}_y$ \eqref{eq:tildeRy} in this section.

\subsection{Solving the bubble equations}
\label{sec:BEsolving}

The bubble equations \eqref{eq:RegRyconnected} are complicated multivariate polynomials of order $N$.  It is difficult to obtain analytic solutions for the system.  However, in the large $N$ limit, numerical computations and analytic approximations can be found. We detail our methods for solving the equations in this limit in the appendix \ref{App:BEanalysis}. 

With odd number of rods, the configuration has a $\mathbb{Z}_2$ symmetry around the middle $(N+1)^\text{th}$ rod and we can fix $M_{2N+2-i}=M_i$ for $i=1,..,N$. We find that the rods have mostly the same size with $\cO(N^{-1})$ corrections except the first and last one. We found that\footnote{More precisely, the ratio between the first rod $M_1$ and the others, $M_i$, $x=\frac{M_1}{M_i}$ is given by the equation $\frac{\Gamma\left[ \frac{1+x}{2}\right]}{\Gamma\left[ \frac{x}{2}\right]} = \frac{2^{1/6}\,\sqrt{e}}{\cA^6}\,.$ }
\begin{equation}
M_i \sim \kappa N^{-\frac{3}{4}} \,\widetilde{R}_y \,,\qquad M_1 \= M_{2N+1} \approx 0.655\,M_i\,,\qquad i=2,3,\ldots,2N\,,
\label{eq:SolutionBE}
\end{equation}
where $\kappa \equi \frac{\cA^3}{2^\frac{25}{12}\,e^{\frac{1}{4}}}  \approx 0.388$ and $\cA$ is the Glaisher-Kindelin constant $\cA \approx 1.282$. Each bubble is then microscopic as the rod lengths scale as $N^{-\frac{3}{4}} \widetilde{R}_y$.  Note that the size of an individual bubble is not strictly speaking the length of the associated rod.  The physical size can be computed considering the curved background such as ``bubble length''$=\int ds|_\text{rod}$ and ``bubble area''$=\int d^3 x\sqrt{\det g|_\text{rod}}$. Even with those considerations, the size of an individual bubble will scale as $N^{s}\widetilde{R}_y$ with $s$ negative.  We discuss more of this around \eqref{eq:AreaBubble}.  


\subsection{General properties and conserved charges}

In the four-dimensional framework \eqref{eq:4dFrameworkcharged}, Weyl stars correspond to static Einstein-Maxwell-dilaton solutions with conserved charges \eqref{eq:conservedchargesrod}
\begin{equation}
\begin{split}
\cM&\, \sim \,N^{\frac{1}{4}}\,\left(3 \coth b_0 \+ 2\coth b_1 -1 \right)\,\frac{\kappa\widetilde{R}_y}{8G_4}\,, \qquad \cQ_m^{(0)} \sim N^\frac{1}{4} \,\frac{3\kappa\widetilde{R}_y}{8\sqrt{\pi\,G_4}\,\sinh b_0},\\
 \cQ^{(1)}_e &\= \cQ_m^{(1)} \, \sim \, N^\frac{1}{4} \,\frac{\kappa \widetilde{R}_y}{8\sqrt{\pi G_4}\,\sinh b_1}\,,
\end{split}
\label{eq:ADMmassWeylStar}
\end{equation}
with the neutral solutions existing in the limit $b_I\to \infty$. Therefore, the conserved charges scale as $N^{\frac{1}{4}} \widetilde{R}_y$ and Weyl stars with large number of smooth microscopic bubbles describe spacetimes with mass (and potentially charges) that is larger than the Kaluza-Klein scales. This is a \emph{novel mechanism} as conical singularities have been so far the only ingredients known to produce large bubbling solutions \cite{Bah:2020ogh,Bah:2020pdz,ourpaper}. 

We have constructed a class of smooth horizonless solutions without conical singularities. The solutions are uniquely determined by their asymptotic data since all the internal $M_i$ are fixed according to $\widetilde{R}_{y_1}= \widetilde{R}_{y_2}=\widetilde{R}_y$ \eqref{eq:tildeRy}, and the number of bubbles $n=2N+1$ and the gauge field parameters $b_I$ are in one-to-one relation with the ADM mass, $\mathcal{M}$, and charges $(\mathcal{Q}_e,\mathcal{Q}_m)$.

However, in order to trust the classical description of the solutions and to neglect quantum effect that may destroy the individual microscopic bubbles, their sizes must be larger than the six-dimensional Planck volume.  Hopefully, the area of individual bubbles are computable analytically \eqref{eq:AreaBubbleAppcharged}, and we must therefore impose 
\begin{equation}
A_{\text{Bu}_i} \equi \int_{y_a,z,\phi} \sqrt{\det g|_{\text{Bu}_i}} \= (2\pi)^2 \,\left(\frac{e^{b_0+b_1}}{4\sinh b_0\,\sinh b_1} \right)^\frac{3}{2}\,M_i\,\widetilde{R}_{y}^2\,\gtrsim\, {\ell_P^{(6)}}^3 .
\label{eq:AreaBubble}
\end{equation}
 If we consider that $ R_{y} = \lambda \,\ell_P^{(4)}$ where $\lambda$ is a dimensionless factor, and use the relation between the four-dimensional and six-dimensional Planck lengths, ${\ell_P^{(6)}}^2 \= R_y \,\ell_P^{(4)}$, we find that  the maximum number of bubbles possible to have a trustworthy solution is
\begin{equation}
N_\text{max} \,\approx\, \lambda^2 \qquad \Rightarrow \qquad \cM_\text{max} \approx \lambda^{\frac{3}{2}}\,m^{(4)}_P\,,
\end{equation}
where $m^{(4)}_P$ is the four-dimensional Planck mass. More concretely, if the extra dimension is $\lambda=\cO(10^{14})$\footnote{This corresponds to KK mass scale at 10 TEV.}   bigger than the four-dimensional Planck length, which is already rather large (of order $10^{-13}$ m), the maximum mass for such solutions is $\cO(10^{21})$ bigger than the Planck mass (of order $10^{13}$ kg). It would be rather optimistic to state that those solutions can live in an astrophysical regime. However, one can bypass this maximal bound by allowing once again conical defects at the bubbles, that are classically resolved into Gibbons-Hawking cycles \cite{Bah:2020ogh,Bah:2020pdz}. For instance, by imposing the same conical defect at each bubble, parametrized by an integer $k\in \mathbb{N}$, the solutions are identical with the rescaling $\widetilde{R}_{y_a} \to k\,\widetilde{R}_{y_a}$, i.e. $\widetilde{R}_{y} \to k\,\widetilde{R}_{y}$. Considering $k$ to be large allows to take arbitrary large mass for the Weyl stars.

\subsection{Profile of Weyl stars}

To best describe Weyl star spacetimes, it is convenient to consider appropriate spherical coordinates centered on the star. For that purpose, we define $M$, the half length of the rod configuration, 
\begin{equation}
M \equiv  z_{2N+1}^+ = \frac{1}{2} \sum_{i=1}^{2N+1} M_i \,\sim\, \kappa\, N^{\frac{1}{4}}\,\widetilde{R}_y,
\label{eq:MdefWeylStar}
\end{equation}
and we change coordinates to
\begin{equation}
\rho = \sqrt{r(r-2M)}\,\sin \theta \,,\qquad z = (r-M) \cos \theta \,,\qquad r\geq 2M \,,~ ~ 0\leq \theta \leq \pi\,.
\label{eq:spheriaccoordWeylstar}
\end{equation}
The base metric, $ds_3^2 = e^{2\nu}(d\rho^2+dz^2)+\rho^2 d\phi^2 $, transforms to
\begin{equation}
ds_3^2 \= e^{2\nu}\,\frac{(r-M)^2-M^2 \cos^2\theta}{r(r-2M)}\,\left(dr^2 + r(r-2M) \,d\theta^2 \right) + r(r-2M) \sin^2\theta \,d\phi^2\,.
\label{eq:basemetricchange}
\end{equation}
The surface of the star, that is the rod configuration, is located at $r=2M$ and $0\leq \theta \leq \pi$, which forms a smooth end to spacetime as already argued. 

\subsubsection{Profile at the surface of the star}

\begin{figure}[h]
\begin{adjustwidth}{-1.2cm}{-1.2cm}
\centering
\includegraphics[width=1.15\textwidth]{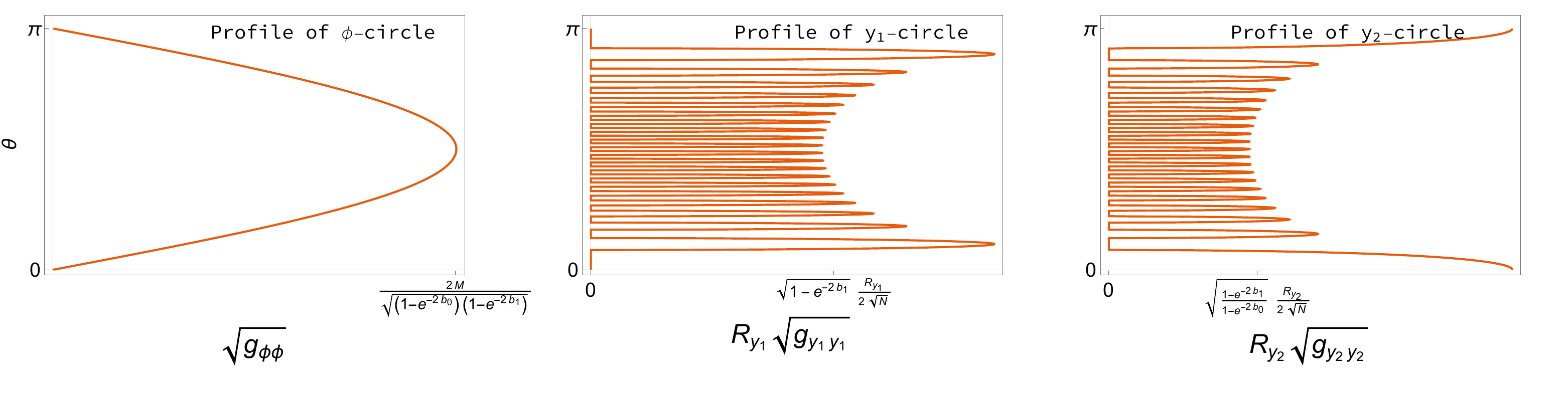}
\end{adjustwidth}
\caption{Sizes of the $\phi$, $y_1$ and $y_2$ circles on the $z$-axis for $N=20$ and $\widetilde{R}_{y_1}=\widetilde{R}_{y_2}=\widetilde{R}_y$. We have written illustrative values on the horizontal axis  that are obtained by computing the values at the middle bubbles while taking the large $N$ limit.}
\label{fig:size41Vac}
\end{figure}

We first study the geometry at the end-to-spacetime locus, $r=2M$. We move along the microscopic bubbles spread across the surface of the Weyl star as $\theta$ varies from $0$ to $\pi$ with $\phi$ fixed. In Fig.\ref{fig:size41Vac}, we have plotted the size of the $\phi$, $y_1$ and $y_2$ circles at $r=2M$ as functions of $\theta$ for a configuration of $41$ microscopic bubbles. The values of each $M_i$ have been obtained numerically and the overall behavior already matches well the large $N$ approximation.

The $y_1$ and $y_2$ circles shrink alternatively as $\theta$ goes from $0$ to $\pi$, and their sizes scale at least as $N^{-\frac{1}{2}}\,\widetilde{R}_y$. Moreover, the bubbles in the middle are more squeezed than the ones at the periphery. The $\phi$-circle is much bigger and scale like $M$ which illustrates that the microscopic bubbles indeed have made the $\phi$-circle to blow up in size. In the large $N$ limit the radius of the $\phi$-circle along $\theta$ direction is very similar to the standard sphere given as
\begin{equation}
\sqrt{g_{\phi \phi}} \,|_{r=2M}\,\sim\, \frac{2 M}{\sqrt{(1-e^{-2 b_0})(1-e^{-2 b_1})}} \,\sin \theta \,.
\end{equation}
Note that the $\phi$ and $y_1$ circles behave as $\left( 1-e^{-2b_1}\right)^{\frac{1}{2}}$ according to the gauge field parameters while the $y_2$ circle has an extra $\left( 1-e^{-2b_0}\right)^{-\frac{1}{2}}$ factor. The neutral limit is obtained for $b_I \to \infty$, and does not change the bubbling feature at the surface of the star as already argued.

Plotting $g_{\theta \theta}$ as a function of $\theta$ is not physically meaningful since it is not associated to any particular symmetry of the system.  We can, however, compute the proper distance from the North pole to the South pole and compare with the ``east-to-west'' distance on the equatorial plane of the star. At leading order in $N$, these are given by
\begin{equation}
\begin{split}
\Delta_{\text{N-S}} &\= \int_0^\pi \sqrt{g_{\theta \theta}}\,|_{r=2M}\,d\theta \,\sim\,  \frac{ N^{\frac{1}{2}} \widetilde{R}_y}{\sqrt{\left( 1-e^{-2b_0}\right)\left( 1-e^{-2b_1}\right)}}\= \cO \left( \frac{G_4^2 \cM^2}{\widetilde{R}_y} \right)\,,\\
 \Delta_{\text{E-W}} &\= \int_0^\pi \sqrt{g_{\phi \phi}}\,|_{r=2M}\,d\phi  \,\sim\, \frac{ N^\frac{1}{4} \widetilde{R}_y}{\sqrt{\left( 1-e^{-2b_0}\right)\left( 1-e^{-2b_1}\right)}}  \= \cO \left( G_4 \cM \right)\,.
\end{split}
\end{equation}
Therefore, the topological S$^2$ at the surface of the star is highly squashed by a factor of $\sqrt{N}$. More precisely, it is a kind of ellipsoid where the proper length from the North pole to the South pole is much larger than the proper length between two opposite point in the equatorial plane, $\Delta_{\text{N-S}} \approx \frac{G_4\cM}{\widetilde{R}_y}\, \Delta_{\text{E-W}} $ (see Fig.\ref{fig:WeylStar}). Note that the physical distance between the North and South pole corresponds to the physical distance between the first and the last microscopic bubbles in the configuration. However, this distance from the point of view of the base is given by $z_{2N+1}^+-z_1^- \sim  G_4 \cM \sim \Delta_{\text{E-W}} $. The physical distance in six dimensions is therefore magnified due to the warp factors.

As a summary, the surface of Weyl stars forms a very special and axisymmetric end-to-spacetime locus with a squashed topological S$^2$.  The T$^2$ has always one of its circle shrinking on the surface defining a chain of axisymmetric bubbles as depicted schematically in Fig.\ref{fig:WeylStar}.

\subsubsection{Spacetime away from the surface}
\label{sec:outofstar}

It is quite difficult to get a good idea of how the solutions look because of the complicated forms of the warp factors \eqref{eq:WarpFactorsVacMultiple} and \eqref{eq:WarpFactorsVacMultipleCharged}. However, now that we have argued that all the rods are very small at large $N$,
\begin{equation}
M_i \= \cO\left( N^{-\frac{3}{4}}\right) \quad \Rightarrow \quad z_i^\pm \= \text{cst}_N \+ \cO \left(\frac{i}{N^{\frac{3}{4}}}\right)\,,
\end{equation}
one can make use of the Riemann sum approximation to greatly simplify the form of the metric, i.e. For differential functions $f$ and $g$, 
\begin{equation}
\sum_{i=1}^N f(x_i)\,\delta x \,\sim\, \int_{x_1}^{x_N} f(x) \,dx \quad \Leftrightarrow \quad \prod_{i=1}^N g(x_i) \,\sim \, \exp \left[\frac{1}{\delta x}\,\int_{x_1}^{x_N} \log g(x) \,dx \right] 
\label{eq:IntApprox}
\end{equation}
if $f=\log g$ is slowly varying in between each segment of size $\delta x$. In our present case, $x_i \= i\,N^{-3/4}$, $\delta x \= N^{-3/4}$, and $f$ will also be a function of the coordinates $(r,\theta)$. 

The arguments of the warp factors, \eqref{eq:WarpFactorsVacMultiple} and \eqref{eq:WarpFactorsVacMultipleCharged}, do indeed vary slowly in segments of size $\delta x \= N^{-3/4}$ outside the rods, i.e. outside the surface of the Weyl star $r\neq 2M$, and vary rapidly on the rods as can be seen in Fig.\ref{fig:size41Vac}. Our procedure is then to apply the approximation to replace each product by an integral for $r>2M$. The integrals are generally derivable and give complicated expressions which we further simplify by considering $N$ large. Thus, there are two stages of approximations for which it is quite difficult to estimate precisely when they fail. We detail our analysis in the appendix \ref{App:RSanalysis}. Taking some margins, we found that the Riemann sum approximation is very accurate for estimating the warp factors \eqref{eq:WarpFactorsVacMultiple} and \eqref{eq:WarpFactorsVacMultipleCharged} when $r \gtrsim 2M\left( 1 + \cO(N^{-1/2}) \right)$. For example, we have
\begin{equation}
U_1 \= \prod_{i=1}^{N+1} \left( 1- \frac{2M_{2i-1}}{r^{(2i-1)}_+ +r^{(2i-1)}_-+M_{2i-1}} \right)\sim \sqrt{1-\frac{2M}{r}}\,,
 \label{eq:approxU1}
\end{equation} in the region outside the star.

Applying the approximation at large $N$ to each warp factor we find that, for $r \gtrsim 2M\left( 1 + \cO(N^{-1/2}) \right)$, the neutral Weyl stars \eqref{eq:neutralWeylmetric} are indistinguishable from the following static axisymmetric spacetimes
\begin{align}
ds_6^2 \=& -dt^2 \+ \sqrt{1-\frac{2M}{r}} \,\left( dy_1^2 +dy_2^2 \right) \+ \left(\frac{(r-M)^2-M^2 \cos^2\theta}{r(r-2M)} \right)^\frac{1}{4} \,\left(\frac{r \,dr^2}{r-2M} + r^2 \,d\theta^2 \right) \nonumber \\
&\+ r^2 \sin^2\theta \,d\phi^2\,.
\label{eq:ApproxMetMultVac}
\end{align}
The three-charge Weyl stars \eqref{eq:met&GFbubbling} are indistinguishable in regions $r \gtrsim 2M\left( 1 + \cO(N^{-1/2}) \right)$ from
\begin{align}
ds_6^2 \=& \frac{1}{\cZ_1}\left[ -dt^2 \+ \sqrt{1-\frac{2M}{r}} \,dy_1^2 \right]  \+ \frac{\cZ_1}{\cZ_0}\,\sqrt{1-\frac{2M}{r}}\,\left( dy_2 \+ \frac{3M}{2\sinh b_0}\cos\theta \, d\phi \right)^2 \nonumber \\
&\+ \cZ_1 \cZ_0\,\left[\left(\frac{(r-M)^2-M^2 \cos^2\theta}{r(r-2M)} \right)^\frac{1}{4} \,\left(\frac{r \,dr^2}{r-2M} + r^2 \,d\theta^2 \right) \+ r^2 \sin^2\theta \,d\phi^2\right] \,, \nonumber \\
F_3 \= &  \frac{M}{2\,\sinh  b_1} \,\left[\frac{1}{r^2\,\cZ_1^2\sqrt{1-\frac{2M}{r}}}\,dt\wedge dr \wedge dy_1 \- \sin \theta \,d\theta \wedge d\phi \wedge dy_2 \right]\,.
\label{eq:ApproxMetMultCharged}
\end{align}
where the $\mathcal{Z}$ functions greatly simplify to
\begin{equation}
 \cZ_0 \equi 1\- \frac{\coth b_0-1}{2} \left(\left(1-\frac{2M}{r}\right)^\frac{3}{2}-1\right)\,,\quad \cZ_1 \equi 1\- \frac{\coth b_1-1}{2} \left(\sqrt{1-\frac{2M}{r}}-1\right). \nonumber
\end{equation}

Remarkably, these axisymmetric systems are exact solutions to Einstein equations with naked curvature singularity at $r=2M$. We refer to them as the \emph{bag spacetimes}. Weyl stars are indistinguishable from bag spacetimes for $r \gtrsim 2M + \cO(\text{bubble size})$,\footnote{We have replaced $N^{-1/2}M \sim N^{-1/4} \widetilde{R}_y$ by the cubic root of the area of each microscopic bubble \eqref{eq:AreaBubble} which we considered as the ``bubble size''.} but as soon as we consider $r\sim 2M$ the smooth chain of microscopic bubbles starts to be manifest and ends the spacetime smoothly as we already described.  \emph{Weyl stars offer a non-trivial classical resolution of the singularity by blowing up bubbles at its vicinity.}

The size of the microscopic bubbles \eqref{eq:AreaBubble} behaves like $\widetilde{R}_y^2\,M^{-1}$, and the ADM mass, $\cM$ \eqref{eq:ADMmassWeylStar}, is proportional to $M$. Therefore, \emph{Weyl stars scale more and more towards the singular solutions as the mass gets large}, i.e. the singular solution becomes more indistinguishable from Weyl stars in the large mass limit. \emph{We have the first examples without the help of supersymmetry of scaling smooth solutions that resemble a singular geometry up to a small scale above its singularity and resolve it into smooth non-trivial topologies}. Even if the singular solution is not a black hole, and doing a similar construction at the vicinity of a black hole horizon is a challenge, this illustrative example shows the strength of the resolution of singularity by microscopic non-trivial smooth topologies in a non-BPS regime.

If the bag spacetimes, \eqref{eq:ApproxMetMultVac} or \eqref{eq:ApproxMetMultCharged}, do not strictly match a black hole metric, it has very interesting properties on its own. The singularity that the Weyl star resolves at $r=2M$ is a naked curvature singularity since several metric coefficients either blow up or vanish. In particular, $g_{\theta \theta}$ and then the area of the S$^2$ diverge. More precisely, the S$^2$ of \eqref{eq:ApproxMetMultCharged} has the expected growth with $r$ at large distances, however, in regions very close to the singularity, the spheres also grow with decreasing $r$.  This implies a minimal surface with respect to the coordinate $r$:
\begin{equation}
\text{Area}_{S^2} \equi \int_{S^2} \sqrt{g_{\theta\theta}g_{\phi\phi}} \underset{r \gg 2M}{\sim} 4 \pi r\,,\qquad \text{Area}_{S^2} \underset{r \sim 2M}{=}  \cO\left( \frac{1}{(r-2M)^{\frac{1}{8}}} \right).
\end{equation}

\begin{figure}[h]
\begin{adjustwidth}{-1.2cm}{-1.2cm}
\centering
\includegraphics[width=1.05\textwidth]{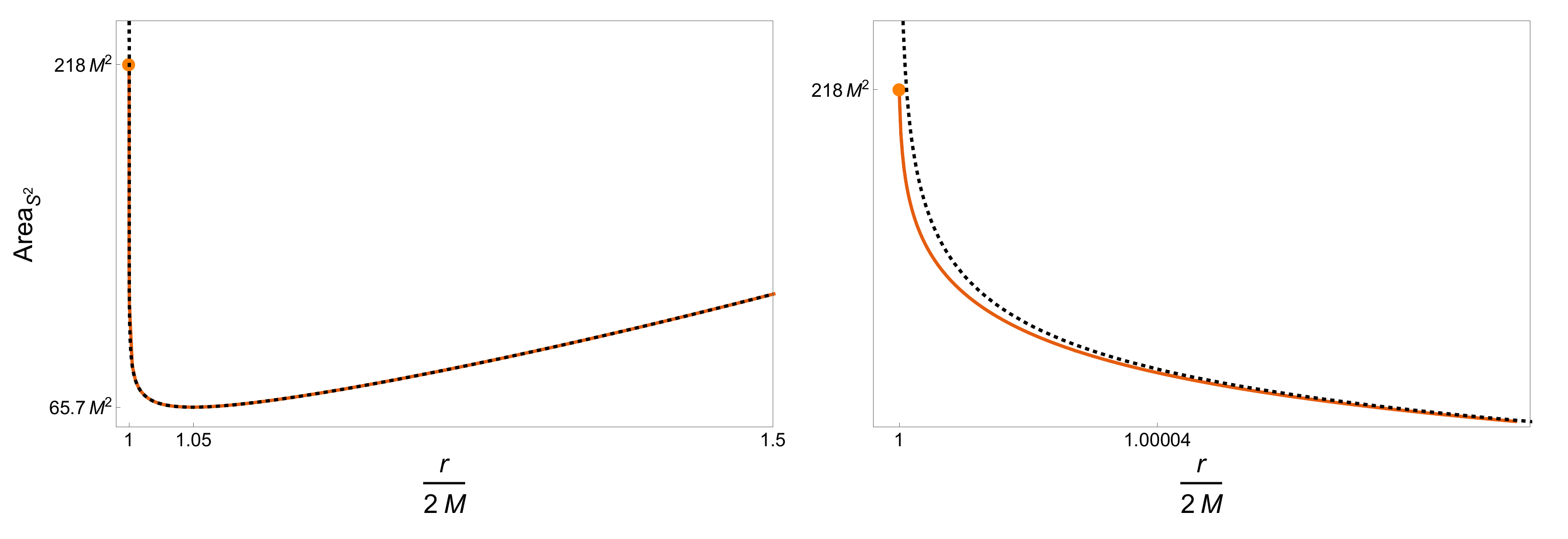}
\end{adjustwidth}
\caption{Area of the S$^2$ as a function of $r/2M$ of a neutral 41-bubble Weyl star (orange thick line) and its approximate bag spacetime away from its surface (black dashed line). At $r=2M$ the area for the Weyl star is finite, $\text{Area}_{S_2}  \approx 218 M^2$, but diverges for the bag geometry. Otherwise, the two results are indistinguishable above $r > 1.00004 \times 2M$, that is very close to the surface of the star even if $N=20$.}
\label{fig:areaS2}
\end{figure}
\begin{figure}[h]
\centering
\includegraphics[width=0.8\textwidth]{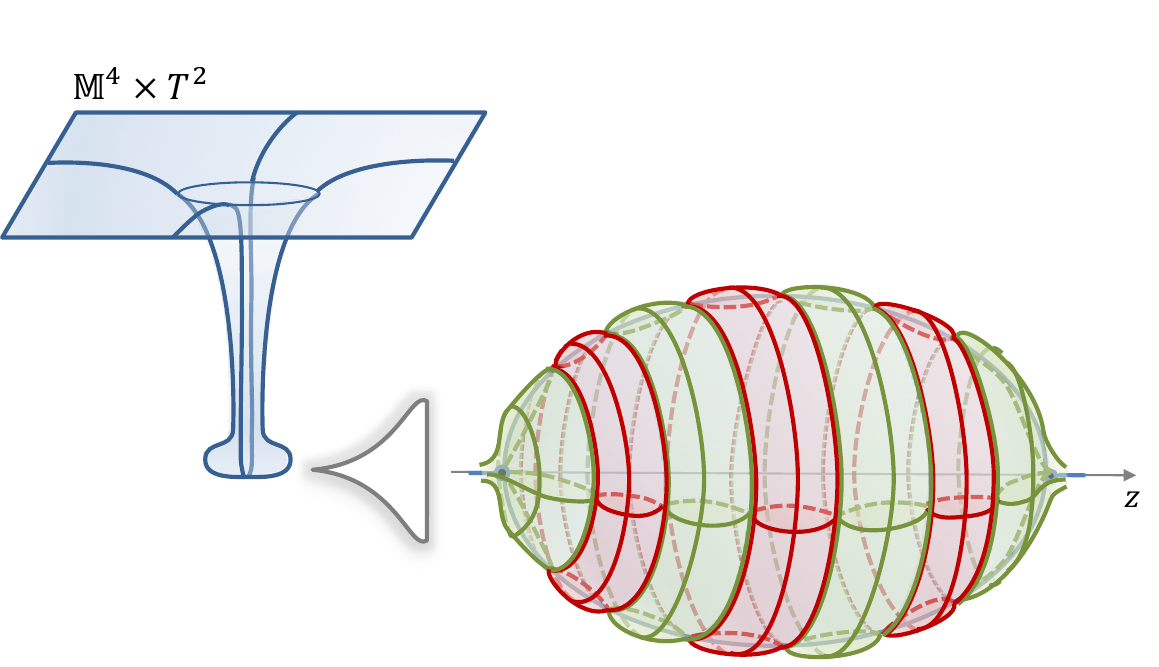}
\caption{Schematic description of a Weyl star spacetime. The T$^2$ circles are colored in red and green. The S$^2$ at the surface of the star is recovered by axisymmetric microscopic bubbles (7 in the figure) where one of the T$^2$ circles shrink smoothly.}
\label{fig:WeylStar}
\end{figure}

We have plotted the area of the S$^2$ for a neutral Weyl star of 41 microscopic bubbles (orange thick line) and for its corresponding bag metric \eqref{eq:ApproxMetMultVac} (black dashed line) in Fig.\ref{fig:areaS2}. The Weyl star resolves the divergence and the area converges to a finite value of order $N^\frac{3}{4} \widetilde{R}_y^2 \sim (G_4\cM)^3 \widetilde{R}_y^{-1}$. Moreover, it matches the area obtained from the bag metric \eqref{eq:ApproxMetMultVac} very well until incredibly close to $r\sim 2M$. This is due to the fact that our range, $r \gtrsim 2M\left( 1 + \cO(N^{-1/2}) \right)$ is mostly restricted to $\theta \sim 0$ or $\pi$ and can be significantly improve away from the axis. Thus, the Weyl stars have the same properties, as soon as we are not considering its surface, as the bag metrics \eqref{eq:ApproxMetMultVac} and \eqref{eq:ApproxMetMultCharged}. \emph{Therefore, a Weyl star also has a minimal S$^2$ and suddenly opens up to a larger space until we reach its surface spread by microscopic bubbles} (see Fig.\ref{fig:WeylStar}). We call such an atypical characteristic \emph{a bubble bag end}.

The minimal surface of the Weyl star is sufficiently away from its surface to be well-captured by the approximated bag metrics, \eqref{eq:ApproxMetMultVac} and \eqref{eq:ApproxMetMultCharged}. Their S$^2$ area can be derived exactly 
\begin{equation}
\begin{split}
\text{Area}_{S^2} \= \frac{2\pi\,r^2\,\cZ_0 \cZ_1}{\left( r(r-2M)\right)^\frac{1}{8}}\,\int_{-1}^1 \left[(r-M)^2-M^2 u^2 \right]^\frac{1}{8}\,du\,,
\label{eq:areaS2WeylStar}
\end{split}
\end{equation}
which can be also expressed in terms of hypergeometric functions. There is no analytic expression for this minimal surface, however it can be studied with the integral form. For neutral solutions, we find numerically that $\min \left(\text{Area}_{S^2}\right) \approx 65.69 M^2 $ at $\frac{r}{2M} \approx 1.049$. Moreover, even if the S$^2$ at the surface of the Weyl star is highly squashed, one can check that the S$^2$ at the minimal surface is mostly round as $1.04 M \leq \sqrt{g_{\theta \theta}}|_{r \sim 2.1M} \leq 1.31 M$. In other words, before reaching this region and opening up into a very axisymmetric space, Weyl star backgrounds are nothing special and appear almost spherically symmetric.

Finally, we have represented the main profile of a Weyl star in Fig.\ref{fig:WeylStar}. The left-hand side describes the geometry away from its surface, obtained from \eqref{eq:ApproxMetMultVac} and \eqref{eq:ApproxMetMultCharged}. The space opens up near the end-to-spacetime locus like a \emph{bubble bag}. The right-hand side corresponds to the smooth geometry of the T$^2\times$S$^2$ at the surface of the star. The S$^2$ is squashed along the axis and spread by axisymmetric microscopic bubbles where the T$^2$ directions alternately degenerate.

From the perspective of the four-dimensional framework obtained by KK reduction along $(y_1,y_2)$ \eqref{eq:4dFrameworkcharged}, three-charge Weyl stars are given by, for $r \gtrsim 2M\left( 1 + \cO(N^{-1/2}) \right)$,
\begin{align}
ds_4^2 \=& - \frac{\sqrt{1-\frac{2M}{r}}}{\sqrt{\cZ_0}\,\cZ_1}\,dt^2 + \frac{\sqrt{\cZ_0}\,\cZ_1}{\sqrt{1-\frac{2M}{r}}} \Biggl[ \left(\frac{(r-M)^2-M^2 \cos^2\theta}{r(r-2M)} \right)^\frac{1}{4} \,\left(\frac{r dr^2}{r-2M} + r^2 \,d\theta^2 \right) \nonumber\\
& \hspace{5cm}  \+  r^2 \sin^2\theta \,d\phi^2 \Biggl] \,,\nonumber \\
e^{-\sqrt{3}\,\Phi_1} \=& \frac{1-\frac{2M}{r}}{\sqrt{\cZ_0}\,\cZ_1}\,,\qquad e^{-\sqrt{\frac{3}{2}}\,\Phi_2} \=  \frac{\cZ_1 \sqrt{ 1-\frac{2M}{r}}}{\cZ_0} \,,
\end{align}
\begin{align}
F^{(m0)} \=& -\frac{3M}{2\sinh b_0}\,\sin \theta \,d\theta \wedge d\phi \,,\qquad  F^{(m1)} \=  -\frac{M}{2\,\sinh  b_1} \,\sin \theta \,d\theta \wedge d\phi\,,\hspace{2.7cm} \nonumber\\
F^{(e1)} \= &  \frac{M}{2\,\sinh  b_1} \,\frac{1}{r^2\,\cZ_1^2\sqrt{1-\frac{2M}{r}}}\,dt\wedge dr \,,\nonumber
\end{align}
and the neutral solutions are obtained by taking $b_I\to \infty$.

\subsection{Comparison with a black brane}

In this section, we compare the geometry of Weyl stars to the six-dimensional black brane with the same conserved charges. To do this, we will compare with the bag spacetimes \eqref{eq:ApproxMetMultVac} and \eqref{eq:ApproxMetMultCharged}.

\subsubsection*{Neutral solutions}

In terms of conserved charges, the geometry of a neutral Weyl star of four-dimensional mass $\cM$ is given by, for $r\gtrsim 4G_4 \cM\left( 1 \+ \cO \left(\frac{\widetilde{R}_y^2}{G_4^2 \cM^2} \right) \right)$,
\begin{equation}
\begin{split}
ds_6^2 \=& -dt^2 \+ \sqrt{1-\frac{4G_4 \cM}{r}} \,\left( dy_1^2 +dy_2^2 \right)\+ r^2 \sin^2\theta \,d\phi^2 \\
& \+ \left(\frac{\left(r-2G_4 \cM\right)^2-\left(2G_4 \cM \cos\theta \right)^2}{r\left(r-4G_4 \cM\right)} \right)^\frac{1}{4} \,\left(\frac{r \,dr^2}{r-4G_4 \cM} + r^2 \,d\theta^2 \right) \,,
\label{eq:ApproxMetMultVacMass}
\end{split}
\end{equation}
and its four-dimensional KK reduction gives
\begin{align}
ds_4^2 = \sqrt{1-\frac{4G_4 \cM}{r}} \Biggl[&-dt^2 + r^2 \sin^2\theta \,d\phi^2 \label{eq:ApproxMetMultVacMass4d} \\
&+ \left(\frac{(r-2G_4 \cM)^2-\left(2G_4 \cM \cos\theta \right)^2}{r\left(r-4G_4 \cM\right)} \right)^\frac{1}{4} \left(\frac{r \,dr^2}{r-4G_4 \cM} + r^2 d\theta^2 \right) \Biggr]\,,\nonumber 
\end{align}
We consider the neutral two-dimensional brane with the same conserved charges
\begin{equation}
ds_6^2 \= - \left( 1- \frac{2G_4 \cM}{r} \right) \,dt^2 \+ dy_1^2 \+dy_2^2 \+ \frac{r\,dr^2}{r-2G_4 \cM}\+ r^2 \left( d\theta^2 + \sin^2 \theta d\phi^2\right)\,,
\end{equation}
which simply corresponds to a T$^2$ fibration over a Schwarzschild black hole and the horizon is located at $r_\text{S}=2G_4 \cM$. It is tempting to compare the radial distance $r=4G_4 \cM$ where the surface of the Weyl star is to the horizon radius $r_\text{S}=2G_4 \cM$ which is twice smaller. However, nothing indicates that the radial coordinates of the two objects are comparable. A meaningful quantity to compare are the minimum S$^2$ surfaces of both solutions and their corresponding radii. For the neutral black brane, the area is obviously the horizon area, $\text{Area}_{S^2, \,\text{hor}} = 16\pi G_4^2 \,\cM^2$ while for the Weyl star, $\text{min}\left(\text{Area}_{S^2} \right) \approx 262.3 \,G_4^2 \,\cM^2$ \eqref{eq:areaS2WeylStar}. Therefore, the ratio of the radii gives
\begin{equation}
\frac{r_\text{min}}{r_\text{S}}\= \sqrt{\frac{\text{min}\left(\text{Area}_{S^2} \right)}{\text{Area}_{S^2, \,\text{hor}} }} \approx 2.29 \,.
\end{equation}
Thus, before the spacetime of the Weyl star opens up, the minimal S$^2$ has a radius $2.28$ bigger than its Schwarzschild radius, which means that the object is very much compact.

From a four-dimensional perspective the S$^2$ of the Weyl star is pinching off at $r\sim 4G_4 \cM$ unlike the Schwarzschild black hole. However, the full six-dimensional geometry should start to be manifest close to this radius and the ``bubble bag end'' that we have already described starts to be visible. 

Pushing the comparison forward will require more work that we postpone for further projects. It would be interesting to compute the multipole moments of Weyl stars that are a priori non-zero due to the strong axisymmetry unlike the Schwarzschild black hole. Moreover, computing and comparing light geodesics could highlight some trapping properties of Weyl stars. 

\subsubsection*{Three-charge solutions}

A three-charge Weyl star \eqref{eq:ApproxMetMultCharged} must be compared to a three-charge non-extremal static black brane \cite{ourpaper}, commonly known as a D1-D5-KKm black hole in type IIB \cite{Cvetic:1995kv,Cvetic:2011dn}. Because the Weyl star has two equal charges (the D1 and D5 in type IIB), we therefore consider the following black-brane solution
\begin{equation}
\begin{split}
ds_6^2 \=&- \left( 1-\frac{r_\text{S}}{r}\right)\,dt^2 \+ \left( 1-\frac{r_\text{B}}{r}\right)\,dy_1^2\+ \left( 1-\frac{r_\text{C}}{r}\right)^{-1}\,(dy_2+P_0\cos\theta d\phi )^2 \\
&\+ \frac{r (r-r_\text{C})\,dr^2}{(r-r_\text{S})(r-r_\text{B})} \+ r\,(r-r_\text{C}) \,d\Omega_2^2, \\
F_3 \= &\frac{Q_1}{r^2} \, dt\wedge dr\wedge dy_1 \- Q_1 \,\sin\theta \,d\theta \wedge d\phi\wedge dy_2 \,,\qquad H_0\, d\phi \= Q_0 \,\cos\theta \, d\phi\,,
\end{split}
\label{eq:metricBBrecap}
\end{equation}
where we have 
\begin{equation}
Q_1^2 \= r_\text{B}\,r_\text{S}\,,\qquad Q_0^2 \= (r_\text{B}-r_\text{C})\,(r_\text{S}-r_\text{C})\,, \qquad r_\text{S} \geq r_\text{B} \geq r_\text{C}\,.
\end{equation}
And the conserved quantities in four dimensions are given by 
\begin{equation}
\cM \= \frac{1}{4 G_4}\,\left( 2 r_\text{S}+r_\text{B}-r_\text{C} \right)\,,\quad\cQ^{(1)}_e \= \cQ^{(1)}_m \= \frac{Q_1}{\sqrt{16 \pi G_4}}\,,\quad \cQ^{(0)}_m \= \frac{Q_0}{\sqrt{16 \pi G_4}}\,.
\end{equation}
It terms of $M$, the conserved charges of a Weyl star \eqref{eq:ADMmassWeylStar} are given by
\begin{equation}
\begin{split}
\cM&\, \= \,\left(3 \coth b_0 \+ 2\coth b_1 -1 \right)\,\frac{M}{8 G_4}\,, \qquad \cQ_m^{(0)} \= \frac{3\,M}{8 \sqrt{\pi G_4}\,\sinh b_0},\\
 \cQ^{(1)}_e &\= \cQ_m^{(1)} \= \frac{M}{8 \sqrt{\pi G_4}\,\sinh b_1}\,.
\end{split}
\end{equation}
To compare the Weyl star with the black brane at fixed conserved charges, we should invert these expressions to obtain $(r_\text{S},r_\text{B},r_\text{C})$ and $(M,b_0,b_1)$ as functions of the conserved charges. Unfortunately, this leads to cubic polynomials and the final results do not simplify, and we will therefore refer to plots and illustrative examples.

\begin{figure}[h]
\begin{adjustwidth}{-1.2cm}{-1.2cm}
\centering
\includegraphics[width=0.7\textwidth]{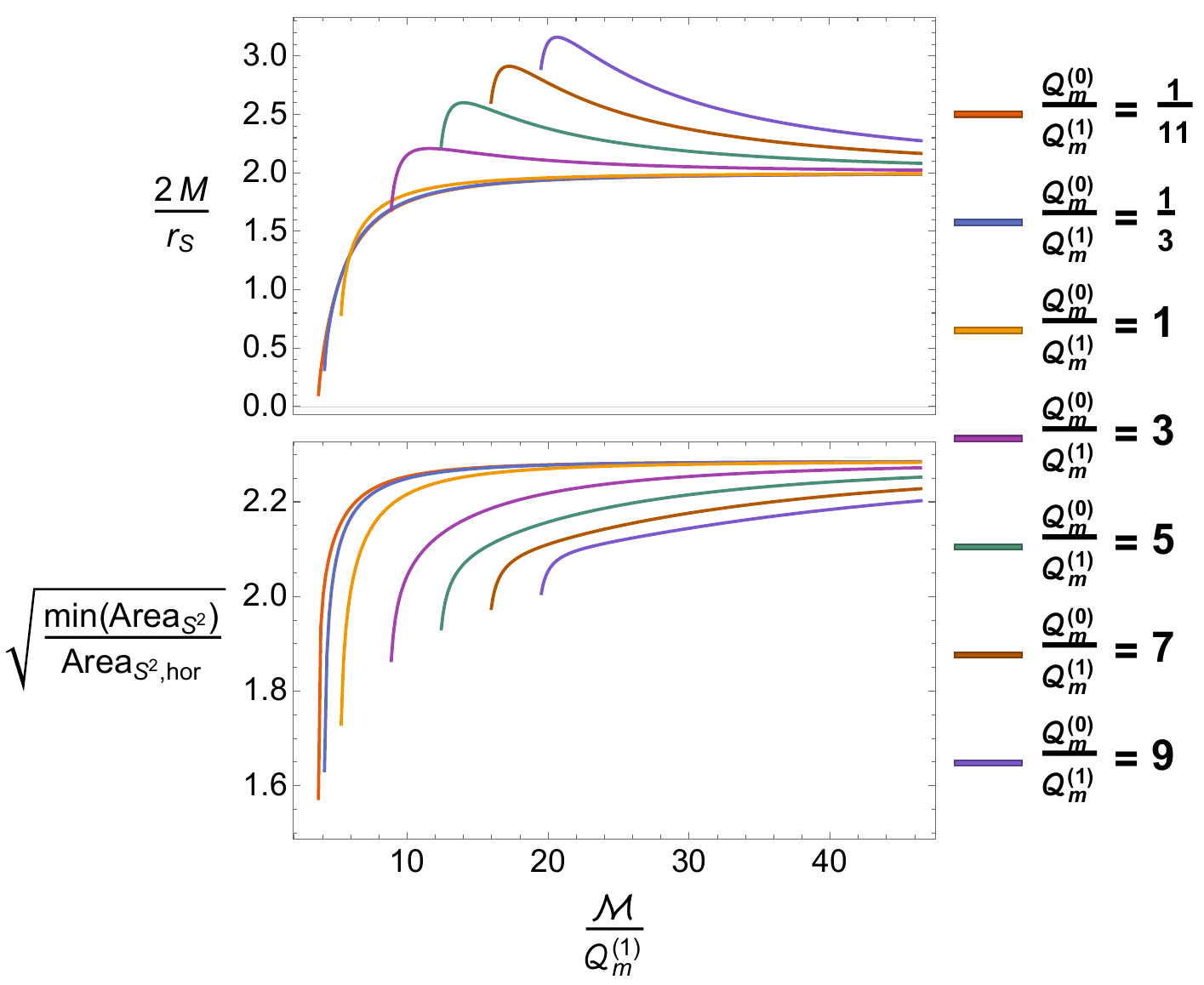}
\end{adjustwidth}
\caption{Plots of $\frac{2M}{r_\text{S}}$ and $\sqrt{\frac{\text{min}\left(\text{Area}_{S^2} \right)}{\text{Area}_{S^2, \,\text{hor}} }}$ as functions of $\frac{\cM}{Q_m^{(1)}}$ for various values of $\frac{\cQ_m^{(0)}}{Q_m^{(1)}}$. The plots are in units of $G_4$ ($G_4=1$). The starting point of each line corresponds to the point in the phase space where the three-charge black brane starts to exist, and then to its extremal limit where $r_\text{S}=r_\text{B}$ in \eqref{eq:metricBBrecap}.}
\label{fig:CompBB}
\end{figure}

In Fig.\ref{fig:CompBB}, the top plot corresponds to the ratio of the surface locus $r=2M$ over the horizon locus $r=r_\text{S}$ for values of mass and charges where both solutions exist. Even if it is interesting to see that in some regime $2M < r_\text{S}$, it does not mean that the Weyl star is more compact than the corresponding black brane. Similar to the neutral solutions, one needs to compare the minimal S$^2$ surfaces of both solutions. The bottom plot of Fig.\ref{fig:CompBB} corresponds to the ratio of the radius of the minimal S$^2$ for the Weyl star, $r_\text{min} = \frac{1}{2\sqrt{\pi}}\sqrt{\text{min}\left(\text{Area}_{S^2} \right)}$ computed from \eqref{eq:areaS2WeylStar}, over the horizon radius of the black brane, $r_\text{H} = \frac{1}{2\sqrt{\pi}}\sqrt{\text{Area}_{S^2,\text{hor}}} \= \sqrt{r_\text{S}(r_\text{S}-r_\text{C})}$. We see that the Weyl star is indeed less compact that the black brane with same charges and mass as expected. Moreover, at a given $\frac{\cQ_m^{(0)}}{Q_m^{(1)}}$, the Weyl star is more compact with respect to the black brane at the extremal point when $r_\text{S}=r_\text{B}$. Numerically, we found that the most compact Weyl star with respect to the black brane of same charges is when $\frac{\cQ_m^{(0)}}{Q_m^{(1)}}$ is approaching zero with the ratio
\begin{equation}
\frac{r_\text{min}}{r_\text{H}}\= \sqrt{\frac{\text{min}\left(\text{Area}_{S^2} \right)}{\text{Area}_{S^2, \,\text{hor}} }} \approx 1.54 \,.
\end{equation}
Finally, all curves converge to the neutral limit as $\cM$ gets large compared to $Q_m^{(1)}=Q_e^{(1)}$ and $Q_m^{(0)}$. Moreover, the ratios of the minimal S$^2$ areas are always smaller than the neutral limit. In other terms, the charged Weyl stars are more compact with respect to the black brane than the neutral configurations. One can wonder then if one can construct even more compact bubbling solutions by turning on other gauge field degrees of freedom (that will allow for instance Chern-Simons interaction). A remarkable scenario will be to have geometries that can be as compact as the equivalent black brane, in the same way that some BPS bubble geometries can be scaling and resemble its corresponding extremal black brane until very close to the horizon. 

As for the neutral Weyl stars, it will be interesting to push the comparison forward through the computation of multipole moments, light geodesics, etc. 

\section{Bag spacetimes and their bubbly resolution}
\label{sec:bagspacetimes}

In the previous section, we highlighted the presence of bag spacetimes which are static and axisymmetric solutions of our six-dimensional action \eqref{eq:Action6d}. These are singular solutions for which the singularity can be resolved into a chain of a large number of smooth bubbles forming a Weyl star. In this section, we reverse the perspective, and we aim to classify and characterize all bag spacetimes and singularities that can be resolved by Weyl stars.

\subsection{The family of singular bag spacetimes}

The bag spacetime obtained in \eqref{eq:ApproxMetMultCharged} is  axisymmetric and static. Therefore, it should have a description in terms of the six-dimensional Weyl formalism pioneered in \cite{ourpaper} and reviewed in section \ref{sec:Weyl} and appendix \ref{App:Weyl6d}. It is clearly sourced by a unique singular rod, and should be characterized in terms of three harmonic functions, $\log W_0$, $L_0$ and $L_1$, \eqref{eq:harmonicfuncGen}, and the weights $(G,P^{(0)},P^{(1)})$ of the source. These singular solutions are not part of the classification of physical Weyl solutions, given in \eqref{eq:BlackBraneGicharged}, \eqref{eq:1TSGi} and \eqref{eq:2TSGi}. However, we now know that these singularities can be classically resolved by  chains of large number of smooth bubbles, and it is therefore relevant to consider them as physical Weyl solutions.

For simplicity, we consider spacetimes sourced by a single rod only, and the generalization to multiple sources on a line is straightforward with the Weyl formalism. We define $2M$ as the length of the rod and the following weights in the harmonic functions:
\begin{equation}
G \=P^{(1)} \= \frac{D}{2} \,,\qquad P^{(0)} \= 1-\frac{D}{2}\,,\qquad 0<D<1.
\end{equation}
The metric and gauge field are given by the generic ansatz \eqref{eq:WeylSol2circle}, and the warp factors and gauge potentials are derived from \eqref{eq:WarpfactorsRods6dApp} such that
\begin{align}
W_0 &= f^{-\frac{D}{2}}, \qquad Z_0 = \frac{e^{b_0} \,f^{-1+\frac{D}{2}}-e^{-b_0}\, f^{1-\frac{D}{2}}}{2\sinh b_0}, \qquad H_0  \= \frac{2-D}{2\sinh b_0} \left(r_- - r_+ \right), \nonumber \\
 Z_1 &= \frac{e^{b_1} \,f^{-\frac{D}{2}}-e^{-b_1}\, f^{\frac{D}{2}}}{2\sinh b_1}, \qquad  H_1 = \frac{D}{2\sinh b_1} \left(r_- - r_+ \right), \\
 T_1 &= -\frac{\sqrt{1+\sinh^2 b_1\, Z_1^2}}{Z_1},\qquad e^{2\nu} = \left(\frac{(r_++r_-)^2-4M^2}{4r_+ r_-} \right)^{1-D(1-D)}, \nonumber
\end{align}
where we have defined
\begin{equation}
r_\pm \equi \sqrt{\rho^2 + (z\mp M)^2}, \qquad f \equi 1-\frac{4M}{r_++r_-+2M},
\end{equation}
It is more convenient to use spherical coordinates, $(r,\theta,\phi)$, centered on the single rod by considering the change of variables \eqref{eq:spheriaccoordWeylstar}. The solutions are therefore given by
\begin{align}
ds_6^2 \=& \frac{1}{\cZ_1}\left[ -dt^2 \+ \left(1-\frac{2M}{r}\right)^D \,dy_1^2 \right]  \+ \frac{\cZ_1}{\cZ_0}\,\left(1-\frac{2M}{r}\right)^{1-D}\,\left( dy_2 \+ \frac{(2-D)M}{\sinh b_0}\cos\theta \, d\phi \right)^2 \nonumber \\
&\+ \cZ_1 \cZ_0\,\left[\left(\frac{(r-M)^2-M^2 \cos^2\theta}{r(r-2M)} \right)^{D(1-D)} \,\left(\frac{r \,dr^2}{r-2M} + r^2 \,d\theta^2 \right) \+ r^2 \sin^2\theta \,d\phi^2\right] \,, \nonumber \\
F_3 \= &  \frac{D\,M}{\sinh  b_1} \,\left[\frac{1}{r^2\,\cZ_1^2\left(1-\frac{2M}{r}\right)^{1-D}}\,dt\wedge dr \wedge dy_1 \- \sin \theta \,d\theta \wedge d\phi \wedge dy_2 \right]\,.
\label{eq:BagCharged}
\end{align}
where the $\mathcal{Z}_I$ functions are 
\begin{equation}
 \cZ_0 \equi 1\- \frac{\coth b_0-1}{2} \left(\left(1-\frac{2M}{r}\right)^{2-D}-1\right)\,,\quad \cZ_1 \equi 1\- \frac{\coth b_1-1}{2} \left(\left(1-\frac{2M}{r}\right)^D-1\right). \nonumber
\end{equation}
To better describe the geometries, it is useful to consider the neutral solutions obtained by sending $b_I \to \infty$, which have the same properties as the charged solutions without the complexity due to the flux:
\begin{align}
ds_6^2 \=& -dt^2 \+ \left(1-\frac{2M}{r}\right)^D \,dy_1^2 \+ \left(1-\frac{2M}{r}\right)^{1-D}\,dy_2^2 \label{eq:BagNeutral} \\
&\+ \left(\frac{(r-M)^2-M^2 \cos^2\theta}{r(r-2M)} \right)^{D(1-D)} \,\left(\frac{r \,dr^2}{r-2M} + r^2 \,d\theta^2 \right) \+ r^2 \sin^2\theta \,d\phi^2\,, \nonumber \
\end{align}
The specific bag spacetime highlighted in the previous section is obtained by taking $D=\frac{1}{2}$ \eqref{eq:ApproxMetMultCharged}. More generically, we have a  one-parameter family of bag solutions with $0<D<1$. All solutions have similar properties to the bag spacetimes with $D=\frac{1}{2}$. They are regular for $r>2M$. However, the T$^2$ is degenerating while the size of the S$^2$ is diverging at $r=2M$. Moreover, as $D$ gets closer to $0$, the solutions converge towards a single-center species-2 bubble while, as $D$ gets closer to $1$, the solutions approach a single-center species-1 bubble. In between, we have an overlay of both sources where the parameter $D$ dials their relative contributions. At the level of the metric, $D$ controls how fast the $y_1$-circle is shrinking compared to the $y_2$-circle. 

From the perspective of the four-dimensional framework obtained by KK reduction along $(y_1,y_2)$ \eqref{eq:4dFrameworkcharged}, bag spacetimes are given by
\begin{align}
ds_4^2 \=& - \frac{\sqrt{1-\frac{2M}{r}}}{\sqrt{\cZ_0}\,\cZ_1}\,dt^2 + \frac{\sqrt{\cZ_0}\,\cZ_1}{\sqrt{1-\frac{2M}{r}}} \Biggl[ \left(\frac{(r-M)^2-M^2 \cos^2\theta}{r(r-2M)} \right)^{D(1-D)} \,\left(\frac{r dr^2}{r-2M} + r^2 \,d\theta^2 \right) \nonumber\\
& \hspace{5cm}  \+  r^2 \sin^2\theta \,d\phi^2 \Biggl] \,,\nonumber \\
e^{-\sqrt{3}\,\Phi_1} \=& \frac{\left(1-\frac{2M}{r}\right)^{\frac{1}{2}+D}}{\sqrt{\cZ_0}\,\cZ_1}\,,\qquad e^{-\sqrt{\frac{3}{2}}\,\Phi_2} \=  \frac{\cZ_1 \left( 1-\frac{2M}{r}\right)^{1-D}}{\cZ_0} \,,\label{eq:bagspacetimeGen}\\
F^{(m0)} \=& -\frac{(2-D)M}{\sinh b_0}\,\sin \theta \,d\theta \wedge d\phi \,,\qquad  F^{(m1)} \=  -\frac{D\,M}{\sinh  b_1} \,\sin \theta \,d\theta \wedge d\phi\,, \nonumber\\
F^{(e1)} \= &  \frac{D\,M}{\sinh  b_1} \,\frac{1}{r^2\,\cZ_1^2\sqrt{1-\frac{2M}{r}}}\,dt\wedge dr \,.\nonumber
\end{align}
Therefore, the four-dimensional ADM mass and the three charges \eqref{eq:AsymptoticExpGen} are given by
\begin{equation}
\begin{split}
\cM& \= \frac{M \left( 2 D \,\coth b_1+(2-D)\,\coth b_0 -D \right)}{4\,G_4}\,,\qquad \cQ_m^{(0)} \= \frac{(2-D)\,M}{\sqrt{16 \pi G_4}\,\sinh b_0}\,,\\
 \cQ_m^{(1)}& \= \cQ_e^{(1)} \= \frac{D\,M}{\sqrt{16 \pi G_4}\,\sinh b_1}.
\end{split}
\end{equation}

In the previous section, we showed that the bag spacetimes with $D=1/2$ can be resolved by specific Weyl star geometries. In the next section, we discuss in general how bag spacetimes with arbitrary $D$ can be resolved in the same way. 

\subsection{Resolution as Weyl stars}

In section \ref{sec:WeylStar}, we have described Weyl stars for which the rescaled extra-dimensional radii, $\widetilde{R}_{y_a}$ \eqref{eq:tildeRy}, are taken to be equal, $\widetilde{R}_{y_1}=\widetilde{R}_{y_2}$. It has allowed to treat the bubble equations in a much simpler way in section \ref{sec:BEsolving}, but it has obviously restricted to specific geometries. In this section, we discuss what we can expect from generic Weyl stars with $\widetilde{R}_{y_1} \neq \widetilde{R}_{y_2}$ and how they resolve the generic class of bag spacetimes described above. We will not go through the details since constructing Weyl stars with $\widetilde{R}_{y_1} \neq \widetilde{R}_{y_2}$ requires a much better understanding of the bubble equations \eqref{eq:RegRyconnected} at large $N$, and we will just sketch the overall picture.

From the bubble equations \eqref{eq:RegRyconnected}, it is clear that considering $\widetilde{R}_{y_1} / \widetilde{R}_{y_2} \neq 1$ will induce a ratio between the lengths of the species-1 bubble rods, $M_{2i-1}$, with respect to the lengths of the species-2 bubble rods, $M_{2i}$. Moreover, as discussed in the appendix \ref{app:BEsolutionsdifferentRy}, we have seen that the distribution of the rod lengths is similar to the case when $\widetilde{R}_{y_1} / \widetilde{R}_{y_2} = 1$, that is all even rods have mostly the same length, with a similar property for the odd rods, except the rods at the periphery. Therefore, for $\widetilde{R}_{y_1} / \widetilde{R}_{y_2}$, the rod lengths are constrained by the bubble equations such that
\begin{equation}
M_{2i+1} \sim m_1, \qquad M_{2i} \sim m_2,\quad i\geq1, \qquad \text{ with }\,\,\frac{m_1}{m_2} \= h\left(N,\frac{\widetilde{R}_{y_1}}{\widetilde{R}_{y_2}}\right),
\end{equation}
where $h$ is a monotically-increasing function of $\frac{\widetilde{R}_{y_1}}{\widetilde{R}_{y_2}}$ with $h(N,1)\sim 1$.

By modifying $\widetilde{R}_{y_1} / \widetilde{R}_{y_2}$ we are therefore changing the \emph{density} of species-1 bubbles compared to the species-2 bubbles in the rod configuration. Indeed, such density is given by 
\begin{equation}
D \= \frac{\sum_{i=1}^{N+1} M_{2i-1}}{\sum_{i=1}^{2N+1} M_{i}} \,\sim\, \frac{m_1}{m_1+m_2} \= \frac{h}{h+1}\,.
\end{equation}
Therefore, having a large $\widetilde{R}_{y_1} / \widetilde{R}_{y_2}$ should induce a large density of species-1 bubbles in the configuration, $D\sim 1$, while having a small $\widetilde{R}_{y_1} / \widetilde{R}_{y_2}$ induces a large density of species-2 bubbles, $D\sim 0$. Beside this difference in the density of bubbles, such Weyl stars have the same properties at the rods as the Weyl stars studied in section \ref{sec:WeylStar}: the spacetime ends off smoothly as a chain of bubbles sitting at origins of $\IR^2$. 

Moreover, one can apply the Riemann sum approximation on the warp factors and gauge potentials $(U_I,\cZ_I,\nu,H_I,T_1)$ \eqref{eq:WarpFactorsVacMultiple} as in section \ref{sec:outofstar}. For $U_1$ for instance, one would find, unlike \eqref{eq:approxU1}, that, at a distance $r$ slightly away from the surface of the star,
\begin{equation}
U_1 \sim \left( 1 - \frac{2M}{r}\right)^{D},
\end{equation}
where $r$ and $M$ are defined as in \eqref{eq:MdefWeylStar} and \eqref{eq:spheriaccoordWeylstar}. More precisely, the overall metric and gauge field are indistinguishable from the bag spacetime described in \eqref{eq:bagspacetimeGen}. 

Therefore, we expect a generic bag spacetime and its singularity \eqref{eq:bagspacetimeGen} to be resolved into a Weyl star, that is a chain of smooth species-1 and species-2 bubbles. The parameter $M$ is related to half the length of the configuration while $D$ is remarkably related to the density of species-1 bubble in the chain. This is consistent with the fact that as $D$ approaches 1, the bag spacetimes \eqref{eq:bagspacetimeGen} converge towards a single-center species-1 bubble.

\section{Discussion}

In this paper, we have shown that interesting solutions can be constructed from the Weyl formalism assuming at least two extra dimensions in addition to the four-dimensional spacetime. In this framework, we have constructed Weyl stars that are smooth static geometries without a horizon, composed of a large number of KK bubbles that can be wrapped by fluxes. These geometries have interesting features with a compact minimal S$^2$ surface that opens to a non-trivial solitonic structure. To that respect we refer them as bubble bag end spacetimes.

An important question to immediately address for the potential astrophysical relevance of such structures is about their stability. Several arguments have been given in \cite{ourpaper}. Because vacuum KK bubbles suffer from a instability that forces them to expand \cite{Witten:1981gj}, we expect that the neutral Weyl stars suffer from the same kind of instability. Indeed, even if the vacuum bubbles in the middle of the chain should be squeezed enough to be prevented from expanding, nothing counterbalances the expansion of the two bubbles at the extremity of the chain. However, it has been argued in \cite{Stotyn:2011tv} that KK bubbles can be stabilized by adding fluxes. Therefore, we hope that the charged Weyl stars are stable and potentially viable astrophysical solutions. To go further and evaluate the possible formation of such structures, one should analyze their thermodynamic properties in the manner of \cite{Gibbons:1978ji,Gross:1982cv,York:1986it,Brown:2014rka}. This is particularly important to better understand the stability properties of Weyl stars.

It is important to describe the gravitational footprints of Weyl stars and bag spacetimes. First, we would like to calculate the geodesics in such backgrounds. It would be particularly interesting to study how probe particles or light can be trapped in the bag region and to image their shadows from an asymptotic observer in the manner of \cite{Bacchini:2021fig}. Secondly, one can think of deriving multipole moments of Weyl stars (such as it has been done for families of BPS and non-BPS horizonless microstate geometries \cite{Bena:2020see,Bena:2020uup,Bianchi:2020miz,Bah:2021jno}), Love number \cite{Cardoso:2017cfl}, quasi-normal modes \cite{Chowdhury:2007jx,Bena:2020yii,Bianchi:2021xpr}, and even gravitational wave emission from Weyl stars. These specific calculations are very relevant to assess how the ringdown signals of Weyl stars with potential echoes may differ from the usual black hole picture or other ultra-compact objects constructed so far.

In this paper, we have also developed a new technique to simplify generic Weyl solutions slightly away from the sources using the Riemann sum approximation, and to obtain simpler backgrounds that solve the Einstein equations by themselves. This technique is not limited to our six-dimensional construction and can be applied to other settings such as the four-dimensional Israel-Khan solutions \cite{Israel1964}, which consist of static black holes on a line, or the five-dimensional black hole-bubble chains \cite{Elvang:2002br}. This would lead to other interesting new bag spacetimes that have not been considered in the past as valid GR solutions because of their naked singularity. However, this singularity is resolved by the sources that constitute the complete Weyl solutions, in the same way that the Weyl stars resolved the singularity of the bag spacetimes constructed in this paper. 

An advantage of our Weyl stars over other toy models of ultra-compact objects like gravastars \cite{Mazur:2001fv} or boson stars \cite{Schunck:2003kk} is that they admit a well-defined UV description in a quantum theory of gravity. Indeed, as described in section \ref{sec:TypeIIB}, they correspond to D1-D5-KKm solutions in type IIB string theory. More precisely, Weyl stars consist of a chain of D1-D5-KKm bubbles and non-BPS KKm bubbles. It would be interesting to analyze the solutions from a more top-down perspective and to better understand their microscopic origin as brane bound states. Furthermore, in this context, the Weyl ansatz and the system of linear equations detailed in section \ref{sec:Weyl} constitute a new example of closed-form ``floating brane'' solutions in the non-BPS regime. The ansatz encompasses the static D1-D5-KKm BPS floating brane ansatz but allows for a departure into the non-BPS regime while retaining the possibility of adding linear brane sources. This is a very non-trivial mechanism as BPS floating brane solutions make extensive use of supersymmetry to keep the brane sources away from each other by imposing a balance between the gravitational and electric interactions. In the present ansatz, the branes can still ``float'' and are kept apart by pure topology, i.e. by KK bubbles. In a future article, we will discuss in more detail, from a top-down point of view, how the Weyl formalism allows to derive a non-BPS floating brane ansatz in M-theory \cite{upcomingour}. 

As a final comment, we would like to draw a parallel between our solutions and BPS microstate geometries \cite{Bena:2004de,Bena:2006kb,Bena:2007qc,Bena:2007kg,Bena:2015bea,Heidmann:2017cxt,Bena:2017fvm} and address the question of whether Weyl stars could correspond to classical microstate geometries of non-extreme D1-D5-KKm black holes. Weyl stars share the same fundamental properties that have been theorized as the only viable mechanism that allows sustainable smooth microstructure in a region where a horizon should form from a GR perspective \cite{Gibbons:2013tqa}: topological cycles wrapped by fluxes. However, our solutions lack Chern-Simons interactions and are clearly not compact enough to have a microstructure as compact as a horizon. Nevertheless, Weyl stars offer good prototypes of what could be the microstate geometries of non-extremal black holes. One can postulate that such microstate geometries may have a minimal S$^2$ of the same size as the horizon of the non-extreme black hole with the same conserved charges. This may then open up into a large smooth ``bag'' spacetime spread by microscopic bubbles. To move towards this analysis, we would need to add extra degrees of freedom that allow Chern-Simons interactions and ultimately rotation to our current Weyl ansatz. This will allow us to derive the most generic non-BPS floating brane ansatz.

\begin{figure}[h]
\centering
\includegraphics[width=0.16\textwidth]{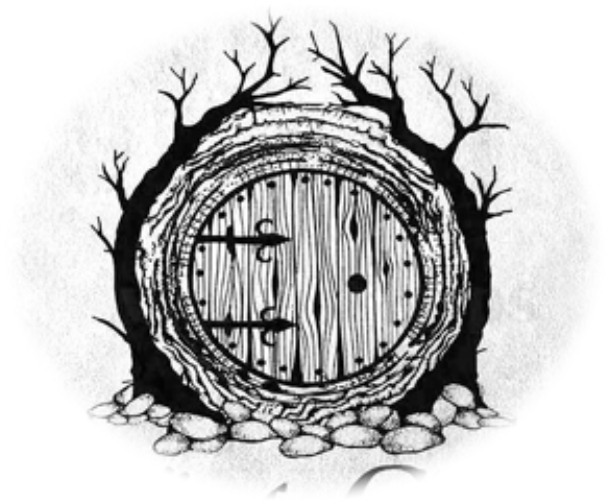}
\end{figure}

\vspace{-0.75cm}
\section*{Acknowledgments}
We thank David Turton for interesting comments and stimulating discussions. The work of IB and PH is supported in part by NSF grant PHY-1820784.

\vspace{1cm}

\newpage
\appendix
\leftline{\LARGE \bf Appendices}

\section{Weyl solutions in six dimensions}
\label{App:Weyl6d}

In this section, we construct static axisymmetric solutions of the six-dimensional Einstein-Maxwell theory \eqref{eq:Action6d}, assuming that $\star F_3 \= F_3$. 

\subsection{Equations of motion}

Using the Weyl's coordinate system, the ansatz that is the most suitable to write down the Einstein-Maxwell equations \eqref{eq:EinsteinMaxEq} is given by 
\begin{align}
ds_{6}^2 = &\frac{1}{Z_1} \left[- W_0\,dt^2 + \frac{ dy_1^2}{W_0} \right] + \frac{Z_1}{Z_0}\, \left(dy_2 +H_0 \,d\phi\right)^2 + Z_0 Z_1\,\left[e^{2\nu} \left(d\rho^2 + dz^2 \right) +\rho^2 d\phi^2\right] ,\nonumber\\
 F_3 \= & d\left[ H_1 \,d\phi \wedge dy_2 \+ T_1 \,dt \wedge dy_1 \right]\, \label{eq:WeylSol2circleApp}
\end{align}
The self-duality of $F_3$ requires
\begin{equation}
dT_1 \= \frac{1}{\rho \,Z_1^2}\,\star_2 dH_1\,,
\end{equation}
where $\star_2$ is the Hodge star operator in the flat $(\rho,z)$ subspace. The Einstein and Maxwell equations can be decomposed into the following layers:
\begin{align}
&\text{\underline{Vacuum layer:}} \quad \cL \log W_0  \= 0\,,\nonumber\\
&\text{\underline{Maxwell layer:}} \quad  \cL \log Z_I \= \- \frac{1}{\rho\, Z_I^2}\,\left[ (\partial_\rho H_I)^2 + (\partial_z H_I)^2 \right] \,,\nonumber\\
&\hspace{2.90cm} \partial_\rho \left( \frac{1}{\rho\,Z_I^2}\,\partial_\rho H_I \right)\+\partial_z \left( \frac{1}{\rho\,Z_I^2}\,\partial_z H_I\right)  \=0\,,\nonumber\\
&\text{\underline{Base layer:}} \nonumber \\
& \partial_z \nu \= \frac{\rho}{2}\, \partial_\rho \log W_0\,\partial_z \log W_0 \+ \rho \,\partial_\rho \log Z_1\,\partial_z \log Z_1 \+ \frac{\rho}{2} \,\partial_\rho \log Z_0\,\partial_z \log Z_0 \nonumber\\
& \hspace{1.3cm} \+ \frac{1}{2 \rho \,Z_0^2} \, \partial_\rho H_0 \partial_z H_0\+ \frac{1}{\rho \,Z_1^2} \,\partial_\rho H_1 \partial_z H_1 \,, \label{eq:EOMWeylApp}\\
& \partial_\rho \nu \= \frac{\rho}{4}\, \left( (\partial_\rho \log W_0)^2- (\partial_z \log W_0 )^2\right)\+ \frac{\rho}{2} \left( ( \partial_\rho \log Z_1)^2- (\partial_z \log Z_1 )^2\right)\nonumber\\
& \hspace{1.3cm} \frac{\rho}{4} \left( (\partial_\rho \log Z_0)^2-(\partial_z \log Z_0)^2\right) + \frac{1}{4 \rho Z_0^2} \left( ( \partial_\rho H_0 )^2-(\partial_z H_0)^2\right)\nonumber \\
& \hspace{1.3cm}+ \frac{1}{2 \rho Z_1^2} \left( (\partial_\rho H_1)^2-( \partial_z H_1 )^2\right)\,, \nonumber
\end{align}
where $\cL$ is the cylindrical Laplacian \eqref{eq:EOM1}. The equations can be treated linearly except the Maxwell layer which is a non-trivial set of non-linear coupled differential equations. In \cite{Bah:2020pdz}, a procedure have been found to extract closed-form solutions. The solutions are determined by three harmonic functions that solve a Laplace equation \eqref{eq:EOM1}, \eqref{eq:DefFi} and \eqref{eq:nuEq}.

\subsection{The BPS limit}
\label{sec:BPSlimitgen}

We consider that the scalars $(Z_I, H_I,T_I)_{I=0,1}$ are given by $ \cG_2^{(I)}$ \eqref{eq:WGF&H}, which leads to $L_I \= Z_I - b_I$. Therefore, the solutions can be better written such that $Z_I$ are harmonic functions
\begin{equation}
\cL \left(\log W_0 \right) \=\cL \left(Z_0\right)\=  \cL \left(Z_1\right) \= 0\,,
\end{equation}
and the gauge potential are determined by
\begin{equation}
\star_3 d (H_0\,d\phi) \= dZ_0\,,\qquad \star_3 d (H_1\,d\phi) \= dZ_1\,,\qquad T_1 \= - \frac{1}{Z_1}\,.
\end{equation}
We recognize the BPS equations of motion one can obtain from supersymmetric static D1-D5-KKm solutions where the D1 and D5 have the same charges. Physical solutions are necessarily sourced by point particles on the $z$-axis with $W_0=1$.\footnote{More precisely, sourcing $Z_I$ by rods or sourcing $\log W_0$ leads to solutions with naked singularities.} With such a choice, $\nu=1$ \eqref{eq:nuEq} and the base is flat $\IR^3$ as expected. Generic solutions are given by
\begin{equation}
Z_I \= \sum_i^n \frac{q_i^{(I)}}{\sqrt{\rho^2+(z-z_i)^2}}\,, \qquad H_I \= \sum_i^n \frac{q_i^{(I)}\,(z-z_i)}{\sqrt{\rho^2+(z-z_i)^2}}\,.
\end{equation}
The sources at $(\rho=0,z=z_i)$ are either static extremal two-dimensional black branes if $q_i^{(1)} \neq 0$ (or D1-D5-KKm black holes when embedded in type IIB) or smooth Gibbons-Hawking centers if $q_i^{(1)}=0$. 

\subsection{Generic three-charge Weyl solutions}
\label{sec:nonBPSgen}

We now move away from the BPS regime and consider that the scalars $(Z_I, H_I,T_I)_{I=0,1}$ are given by $ \cG_1^{(I)}$ \eqref{eq:WGF&H}.\footnote{We could think about mixing the BPS branch with the non-BPS one, by taking $\cG_2^{(0)}$ and $\cG_1^{(1)}$ for instance, but our attempts did not lead to any physical solutions.} Point sources lead to naked singularity, so we source the harmonic function $(\log W_0,L_0,L_1)$ by rods. 

\begin{figure}[h]
\centering
\includegraphics[width=0.23\textwidth]{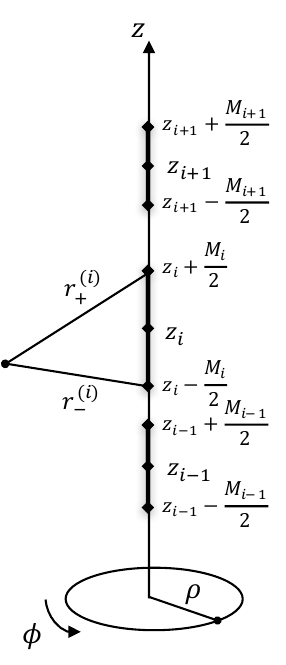}
\caption{Schematic description of axisymmetric rod sources.}
\label{fig:RodsSources}
\end{figure}
\begin{figure}[h]
\centering
\includegraphics[width=0.8\textwidth]{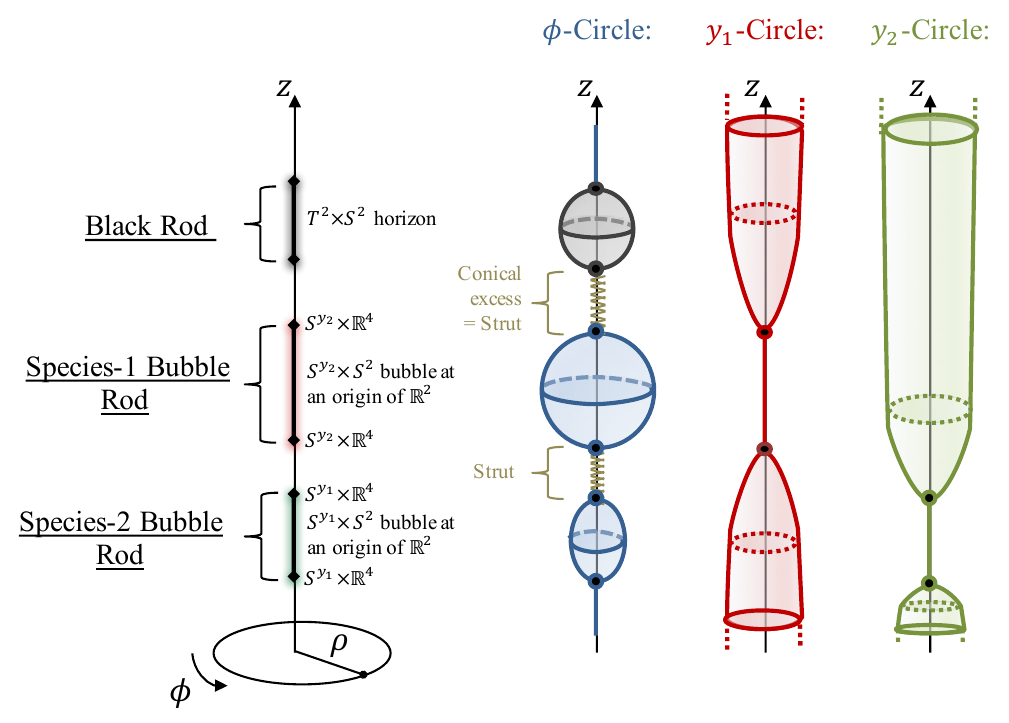}
\caption{Schematic description of the three possible categories of physical rods and the behavior of the circles on the $z$-axis. The black rod sources a two-dimensional static black brane with a T$^2\times$S$^2$ horizon. The species-1 bubble rod corresponds to the degeneracy of the $y_1$-circle inducing a S$^{y_2}\times$S$^2$ bubble and reversely for the species-2 bubble rod. Each section between sources has a conical excess.}
\label{fig:RodCategories}
\end{figure}
\begin{figure}[h]
\centering
\includegraphics[width=0.8\textwidth]{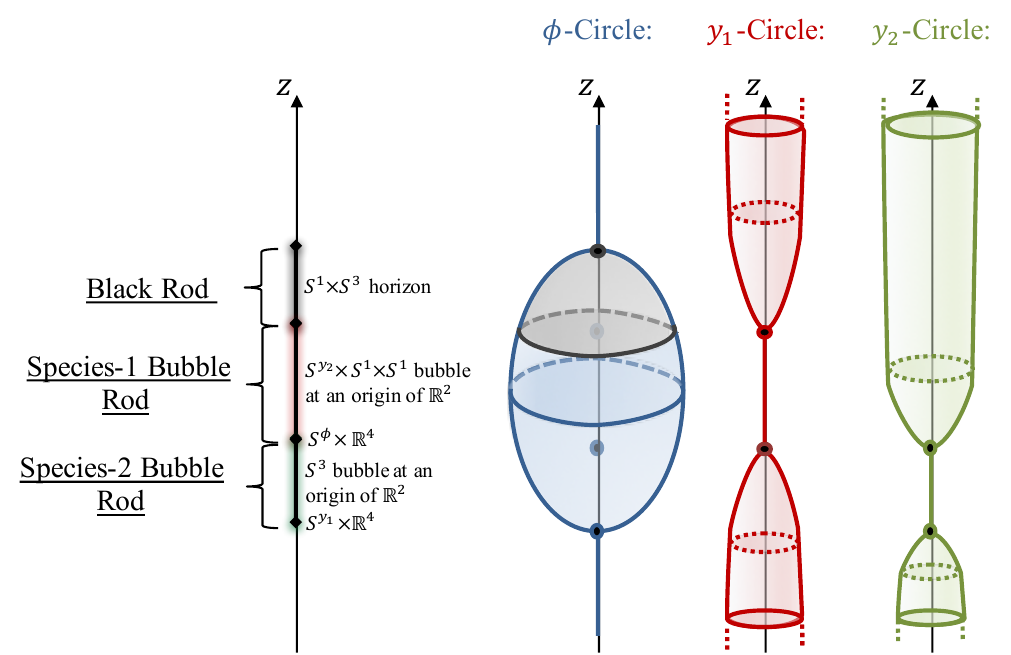}
\caption{Schematic description of connected rods by making the rods of Fig.\ref{fig:RodCategories} to touch. The $\phi$-circle describes now a single bubble without struts. The black rod now describes a black string as the $\phi$-circle degenerates at its North pole and the $y_1$-circle at its South pole. In between the two different bubbles, both the $y_1$ and $y_2$ circle shrinks defining an origin of $\IR^4$.}
\label{fig:TouchingRods}
\end{figure}

 We consider $n$ distinct rods of length $M_i>0$ along the $z$-axis centered on $z=z_i$. Without loss of generality we can order them as $z_i < z_j$ for $i<j$. Our conventions are illustrated in Fig.\ref{fig:RodsSources}. The coordinates of the endpoints of the rods on the $z$-axis are given by
\begin{equation}
z^\pm_i \equi z_i \pm \frac{M_i}{2}\,.
\label{eq:coordinatesEndpoints}
\end{equation}
We define the distances to the endpoints $r_\pm^{(i)}$, the distances $R_\pm^{(i)}$ and the generating functions $E_{\pm \pm}^{(i,j)}$ such as
\begin{equation}
r_\pm^{(i)} \equiv \sqrt{\rho^2 + \left(z-z^\pm_i\right)^2}\,, \quad R_\pm^{(i)} \equiv r_+^{(i)}+r_-^{(i)}\pm M_i\,, \quad E_{\pm \pm}^{(i,j)} \equiv r_\pm^{(i)} r_\pm^{(j)} + \left(z-z_i^{\pm}  \right)\left(z-z_j^{\pm}  \right) +\rho^2,
\label{eq:RpmdefApp}
\end{equation}
The harmonic functions sourced at the rods are given by 
\begin{equation}
\log W_0 =  \sum_{i}^n G_i\,\log \left(  \frac{R_+^{(i)}}{R_-^{(i)}}\right)\,, \qquad L_I \= \sum_{i}^n P_i^{(I)}\,\log \left(  \frac{R_+^{(i)}}{R_-^{(i)}}\right)\,,
\label{eq:HarmFuncApp}
\end{equation}
where $(G_i,P_i^{(0)},P_i^{(1)})$ are weights associated to each rod. The metric warp factors and the gauge potentials, obtained from \eqref{eq:WGF&H} and \eqref{eq:nuEq}, give
\begin{align}
Z_I &\= \frac{1}{2a_I} \left[e^{b_I} \prod_{i=1}^n \left(  \frac{R_+^{(i)}}{R_-^{(i)}}\right)^{a_I P^{(I)}_i}-e^{-b_I} \prod_{i=1}^n \left(  \frac{R_-^{(i)}}{R_+^{(i)}}\right)^{a_I P^{(I)}_i} \right]\,,\qquad W_0 = \prod_{i=1}^n \left(  \frac{R_+^{(i)}}{R_-^{(i)}}\right)^{G_i} \,,\nonumber\\
 e^{2\nu} &\=\prod_{i,j=1}^n\, \left(  \frac{E_{+-}^{(i,j)}E_{-+}^{(i,j)}}{E_{++}^{(i,j)}E_{--}^{(i,j)}}\right)^{\frac{1}{2}\,\alpha_{ij}} \,, \qquad H_0 \= \sum_{i=1}^n P^{(0)}_i \left(r_-^{(i)}-r_+^{(i)} \right)\,, \label{eq:WarpfactorsRods6dApp}\\
H_1 &\=\sum_{i=1}^n P^{(I)}_i \left(r_-^{(i)}-r_+^{(i)} \right)\,,\qquad Z_1\=- a_1 \,\coth \left(\sum_{i=1}^n a_1 P^{(1)}_i \,\log  \frac{R_+^{(i)}}{R_-^{(i)}} +b_1\right),\nonumber
\end{align}
where the exponents $\alpha_{ij}$ are given by
\begin{equation}
\alpha_{ij} \equi G_i G_j \+ a_0^2 \,P^{(0)}_i P_j^{(0)}\+ 2 a_1^2\, P^{(1)}_i P_j^{(1)} \,.
\label{eq:alphaijDef}
\end{equation}
The solutions are asymptotic to T$^2\times\mathbb{M}^4$ if $Z_I  \to 1$, which fixes
\begin{equation}
a_I \= \sinh b_I\,.
\label{eq:AsymptoticCond}
\end{equation}
This implies $b_I\geq 0$. Regularity at the rods fixes the weights $(G_i,P_i^{(0)},P_i^{(1)})$ to three categories of physical sources \cite{ourpaper} (see Fig.\ref{fig:RodCategories} for a schematic description). The exponents $\alpha_{ij}$ \eqref{eq:alphaijDef} simplifies to
\begin{equation}
\alpha_{jk} \= \begin{cases} 
1 \qquad &\text{if the }j^\text{th}\text{ and }k^\text{th}\text{ rods have the same nature.}  \\
\frac{1}{2} \qquad &\text{otherwise,}
\end{cases}
\end{equation}
and it will be convenient to define the following functions of aspect ratios:
\begin{equation}
\begin{split}
d_1 &\equi 1\,,\qquad d_i \equi  \prod_{j=1}^{i-1} \prod_{k=i}^n \left(\dfrac{(z_k^- - z_j^+)(z_k^+ - z_j^-)}{(z_k^+ - z_j^+)(z_k^- - z_j^-)}  \right)^{\alpha_{jk}}\quad \text{when } i=2,\ldots n\,.
\end{split}
\label{eq:dialphaDefcharged}
\end{equation}
The three categories of physical rods are:
\begin{itemize}
\item[•] \underline{Black rod:}  the $i^\text{th}$ rod corresponds either to the horizon of a static three-charge black brane with a T$^2\times$S$^2$ topology or a black string with a S$^1\times$S$^3$ topology\footnote{\label{footnote1}The topology might change whether the rod is connected to other rods (see Fig.\ref{fig:TouchingRods}).} providing that 
\begin{equation}
P^{(0)}_i \= \frac{1}{2\,\sinh b_0} \,,\qquad P^{(1)}_i \= \frac{1}{2\,\sinh b_1} \,,\qquad G_i \= - \frac{1}{2}\,.
\label{eq:BlackBraneGicharged}
\end{equation} 
The surface gravity and the area of the horizon are given by \cite{ourpaper}
\begin{equation}
\kappa_i \equi \frac{\sinh b_1}{e^{b_1}}\,\sqrt{\frac{2\sinh b_0}{e^{b_0}}}\frac{1}{d_i\,M_i}\, \prod_{j\neq i} \left(\frac{z_j^+ - z_i^-}{z_j^- -z_i^-} \right)^{\text{sign}(i-j)\,\alpha_{ij}}\,,\quad A_i \=  \frac{(2\pi)^3}{\kappa_i}\,M_i\,R_{y_1}R_{y_2}\,.
\label{eq:surfaceGrav}
\end{equation}
The gauge fields \eqref{eq:WarpfactorsRods6dApp} are regular at the horizon and carries two magnetic charges (one is coming from the Kaluza-Klein monopole along $y_2$) and an electric charge, that can be derived by integrating the fluxes at the rod. We find
\begin{equation}
\begin{split}
 \cQ_{e\,i}^{(1)}&\={\cQ_{m\,i}^{(1)}} \=  \frac{1}{8\sqrt{\pi G_4}}\,\frac{M_i}{\sinh b_1} \,, \qquad \cQ_{m\,i}^{(0)}  \= \frac{M_i }{8\sqrt{\pi G_4 }\,\sinh b_0}\,,
 \label{eq:ChargesBBcharged}
\end{split}
\end{equation}
where $G_4 = \frac{G_6}{(2\pi)^2 R_{y_1}R_{y_2}}$ is the four-dimensional Newton constant and we use the convention of \cite{Myers:1986un} to derive the charges. Note that the electric and magnetic charges coming from $F_3$ are equal due to the self-duality of $F_3$. In \cite{ourpaper}, we did not assume the self-duality which allowed an arbitrary ratio between these charges.
\item[•] \underline{Species-1 bubble rod:} the $i^\text{th}$ rod corresponds to a static three-charge bubble where the $y_1$-circle shrinks providing that (see Fig.\ref{fig:RodCategories})
\begin{equation}
P^{(0)}_i \= \frac{1}{2\,\sinh b_0} \,,\qquad P^{(1)}_i \= \frac{1}{2\,\sinh b_1} \,,\qquad G_i \=  \frac{1}{2}\,.
\label{eq:1TSGi}
\end{equation} 
The bubble has a S$^1\times$S$^2$ or S$^3$ topology\footnoteref{footnote1}  and the $(\rho,y_1)$ subspace defines an origin of a $\IR^2/\mathbb{Z}_{k_i}$ space if 
\begin{equation}
k_i\,R_{y_1} \=\frac{e^{b_1}}{\sinh b_1}\sqrt{\frac{e^{b_0}}{2\sinh b_0}}\,d_i\,M_i\, \prod_{j\neq i}^n \left(\frac{z_j^+ - z_i^-}{z_j^- -z_i^-} \right)^{\text{sign}(j-i)\,\alpha_{ij}}\,,\qquad k_i \in \mathbb{N}\,.
\label{eq:RegRy1}
\end{equation}
One can also check that the gauge fields are regular, and the bubble carries an electric and two magnetic charges given by the same expression as in \eqref{eq:ChargesBBcharged}. 

Finally, the area of the bubble for a configuration without black rods is remarkably derivable \cite{ourpaper}, and gives
\begin{equation}
A_{\text{B}i} \= \int_{\text{bubble}} \sqrt{\det g|_{\text{bubble}}} \= (2\pi)^2 \, \sqrt{\frac{e^{b_1}}{2 \sinh b_1}}\,k_i\,M_i\,R_{y_1}R_{y_2}\,.
\label{eq:AreaBubbleAppcharged}
\end{equation}
\item[•] \underline{Species-2 bubble rod:} The $i^\text{th}$ rod corresponds to a static one-charge bubble where the $y_2$-circle shrinks providing that (see Fig.\ref{fig:RodCategories})\footnote{It is a one-charge bubble since $P^{(1)}_i=0$ and $F_3$ is then not sourced at the rod.}
\begin{equation}
P^{(0)}_i \= \frac{1}{\sinh b_0} \,,\qquad P^{(1)}_i \= G_i \=  0\,.
\label{eq:2TSGi}
\end{equation} 
The bubble has a S$^1\times$S$^2$ or S$^3$ topology\footnoteref{footnote1} and the $(\rho,y_2)$ subspace defines an origin of a $\IR^2/\mathbb{Z}_{k_i}$ space if 
\begin{equation}
k_i\,R_{y_2} \=\frac{e^{b_0}}{\sinh b_0}\,d_i\,M_i\, \prod_{j\neq i}^n \left(\frac{z_j^+ - z_i^-}{z_j^- -z_i^-} \right)^{\text{sign}(j-i)\,\alpha_{ij}}\,, \qquad k_i \in \mathbb{N}\,.
\label{eq:RegRy2}
\end{equation}
Moreover, the gauge fields are regular at the rod and the bubble carries only a magnetic charge in the KK gauge field $H_0 d\phi$:
\begin{equation}
 \cQ_{e\,i}^{(1)}\={\cQ_{m\,i}^{(1)}} \= 0 \,, \qquad \cQ_{m\,i}^{(0)}  \= \frac{M_i }{4\sqrt{\pi G_4 }\,\sinh b_0}\,.
\end{equation}
The area of the bubble is also derivable when the configuration has no black rods and gives the same expression as \eqref{eq:AreaBubbleAppcharged}
\end{itemize}

If we consider multiple bubbles on the axis, their regularity conditions \eqref{eq:RegRy1} and \eqref{eq:RegRy2} constrain non-trivially the geometry and the size of the rods. These constraints are the equivalent of the \emph{bubble equations} for BPS multicenter solutions \cite{Denef:2000nb,Denef:2002ru,Bates:2003vx}.

Details of the regularity analysis out of the rods but still on the $z$-axis can be found in \cite{ourpaper}. The $\phi$-circle shrinks smoothly as the common cylindrical coordinate degeneracy in the semi-infinite segments above and below the rod configuration. However, any segments that separate two rods are \emph{struts} and have a conical excess of order $d_i<1$ given in \eqref{eq:dialphaDefcharged} (as depicted in Fig.\ref{fig:RodCategories}). If two rods of different nature touch, the strut that separates them disappears and the intersection is free from conical excess (as depicted in  Fig.\ref{fig:TouchingRods}). For instance, the intersection between two touching species-1 and species-2 bubbles is free from conical excess and the local topology correspond to a S$^\phi$ fibration over an origin of an $\IR^4$ with the same potential orbifold defects as at each bubble. 

At this level, there is a large diversity of solutions one can think about. On one side, one can be interested in studying interacting static three-charge black branes either with struts between them as in \cite{Costa:2000kf} or separated by bubbles as in \cite{Elvang:2002br}. The present black branes are richer than the black holes studied in \cite{Costa:2000kf,Elvang:2002br} as they are non-trivially wrapped by fluxes and can be seen as static D1-D5-KKm non-extremal black holes, also called Cvetic-Youm black holes \cite{Cvetic:1995kv,Cvetic:2011dn}. On the other hand, one can be interested in the physics of smooth and strut-free geometries made of connecting species-1 and species-2 bubbles, and this is the subject of this paper. These solutions can be also described from a top-down perspective as smooth D1-D5-KKm floating branes in a non-BPS regime as shown in section \ref{sec:TypeIIB}.

\section{Bubble equations at large $N$}
\label{App:BEanalysis}

In this section, we review our analysis to solve the bubble equations \eqref{eq:RegRyconnected} for a large number of bubbles $n=2N+1$, with the extra assumption that $\widetilde{R}_{y_1} = \widetilde{R}_{y_2} = \widetilde{R}_{y}$ \eqref{eq:tildeRy} and for bubbles that are free from conical defects, $k_i \= 1$.  With these assumptions, one needs to fix all $\{ M_i\}_{i=1,..,2N+1}$ such that
\begin{align}
& 2\,M_1 \prod_{j=1}^{N} \sqrt{1+\frac{M_{2j}}{\sum_{\ell=1}^{2j-1}M_\ell}}\,\prod_{j=1}^{N} 1+\frac{M_{2j+1}}{\sum_{\ell=1}^{2j}M_\ell} \= \widetilde{R}_{y}\,,\nonumber\\
& 2M_i \,\prod_{j=1}^{i-1}\prod_{k=i+1}^{2N+1} \,\left( \left( 1+\frac{M_{j}}{\sum_{\ell=j+1}^{k}M_\ell}\right)\left(1-\frac{M_{j}}{\sum_{\ell=j}^{k-1}M_\ell} \right)\right)^{\alpha_{jk}} \label{eq:RegRyconnectedApp}\\
&\hspace{1cm} \times \prod_{j=1}^{i-1} \left(1+\frac{M_{j}}{\sum_{\ell=j+1}^{i}M_\ell} \right)^{\alpha_{ij}} \prod_{j=i+1}^{2N+1} \left(1+\frac{M_{j}}{\sum_{\ell=i}^{j-1}M_\ell} \right)^{\alpha_{ij}}  \=  \widetilde{R}_{y}\,,\nonumber
\end{align}
where $\alpha_{jk}$ is $1$ if $j$ and $k$ have the same parity or $1/2$ otherwise \eqref{eq:dialphaDefcharged}. These equations are non-trivial multi-variate polynomials of highest degree that scales with $N$. To our knowledge, we cannot solve them explicitly. We will first detail our numerical analysis before our analytic computation at large $N$.

Because the rod configuration is $\mathbb{Z}_2$ symmetric around the middle $(N+1)^\text{th}$ bubble, one can check that the bubble equations are invariant under the change $i \to 2N+2-i$ and then $M_{2N+2-i} = M_i$ for $i=1,..,N$. It is therefore enough to consider only the $N+1$ first bubble equations.

\subsection{Numerical analysis}

\begin{figure}[h]
\begin{adjustwidth}{-1.2cm}{-1.2cm}
\centering
\includegraphics[width=0.75\textwidth]{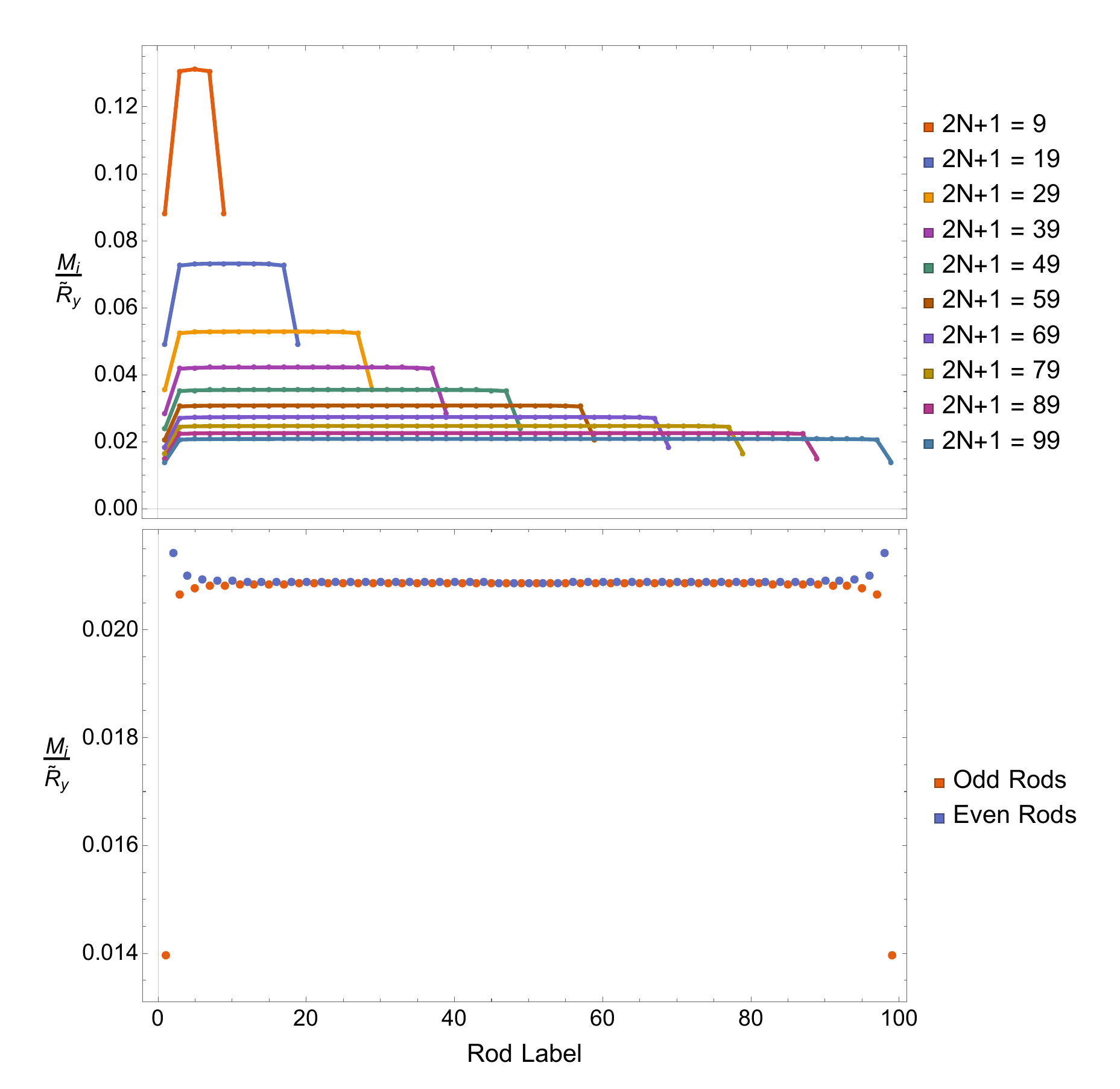}
\end{adjustwidth}
\caption{Plots of the rod lengths $M_i$ as a function of the rod label, $i$, for different rod configurations. The top plot gives the lengths of the odd rods, $M_{2i-1}$, for different total numbers of bubbles, starting with a configuration of 9 bubbles up to a configuration of 99 bubbles. The bottom plot shows the lengths of the even and odd rods for the configuration of 99 bubbles.}
\label{fig:RodSizeNum}
\end{figure}

We have numerically solved the bubble equations for several values of $N$ with a maximal value of $N=49$ (that is a configuration of 99 bubbles). The results are presented in Fig.\ref{fig:RodSizeNum}. First, we have found that the rod lengths, for a given $N$, are mostly equal except the first and last one. Second, the rod lengths are decreasing as a function of $N$. With a numerical fit, we observed that $M_i \propto N^{-\frac{3}{4}}$. Finally, in the bottom plot, we see that the even and odd rod lengths are mostly the same when $\widetilde{R}_{y_1} = \widetilde{R}_{y_2} = \widetilde{R}_{y}$, with a slight difference that the even rods are slightly bigger than the odd ones.

\subsection{Analytic analysis at large $N$}

The numerical analysis suggests that an analytic computation can be performed by assuming that all rod lengths are equal at large $N$ except the first and last one. As a warm-up, we will first derive the bubble equations by considering all rods equal.

\subsubsection*{When all $M_i$ are taken to be equal}

We assume that $ M_i = m,$ for all $i\leq 2N+1$. The bubble equations drastically simplify to
\begin{align}
& \frac{2^{1-N}(1+2N)\sqrt{(2N)!}}{N!}\,m\= \widetilde{R}_{y}\,,\nonumber\\
& 2\sqrt{2}\,m \,\prod_{j=1}^{i-1}\prod_{k=i+1}^{2N+1} \,\left( 1-\frac{1}{(k-j)^2}\right)^{\alpha_{jk}} \, \prod_{j\neq i,i-1}^{2N+1} \left(1+\frac{1}{|i-j|} \right)^{\alpha_{ij}}  \=  \widetilde{R}_{y}\,,\quad i\neq 1\,.\nonumber
\end{align}
By expanding at large $N$, the bubble equations transforms to\footnote{We have taken $j = \left\lfloor \frac{i}{2}\right\rfloor$ in the expansion. }
\begin{equation}
\frac{4 m N^\frac{3}{4} }{\pi^\frac{1}{4}} = \widetilde{R}_{y}\,, \quad  \frac{2^{j(1+2j)} \sqrt{(2j)!}m N^\frac{3}{4} }{\pi^{j+\frac{1}{4}}} \left( \frac{\binom{2j+1}{j}}{(j+2)\,j!}\right)^2 \prod_{k=1}^{j-1} \binom{2k+4}{k+1}^{-2} = \widetilde{R}_{y}, ~ 1\leq j\leq \left\lfloor\frac{N+1}{2}\right\rfloor,
\end{equation}
Obviously, these relations cannot be satisfied since taking all $M_i$ equal is a too strong constraint. However, it is still instructive to see that all equations give
\begin{equation}
\frac{m}{\kappa_j} \, N^\frac{3}{4} \= \widetilde{R}_{y}\,,
\end{equation}
where $\kappa_j$ is a real number in between $[\kappa_1, \kappa_{\left\lfloor\frac{N+1}{2}\right\rfloor}] \= [\frac{\pi^\frac{1}{4}}{4}\,,\, \frac{\cA^3}{2^\frac{25}{12}\,e^{\frac{1}{4}}} ] \approx [0.33\,,\, 0.39]$. Because all $\kappa_j$ are very close to each other, the approximation is almost working. In the next section we will improve the approximation by considering that the first and last rods have a different length.

\subsubsection*{When $M_1$ and $M_{2N+1}$ are different}

\begin{figure}[h]
\begin{adjustwidth}{-1.2cm}{-1.2cm}
\centering
\includegraphics[width=0.75\textwidth]{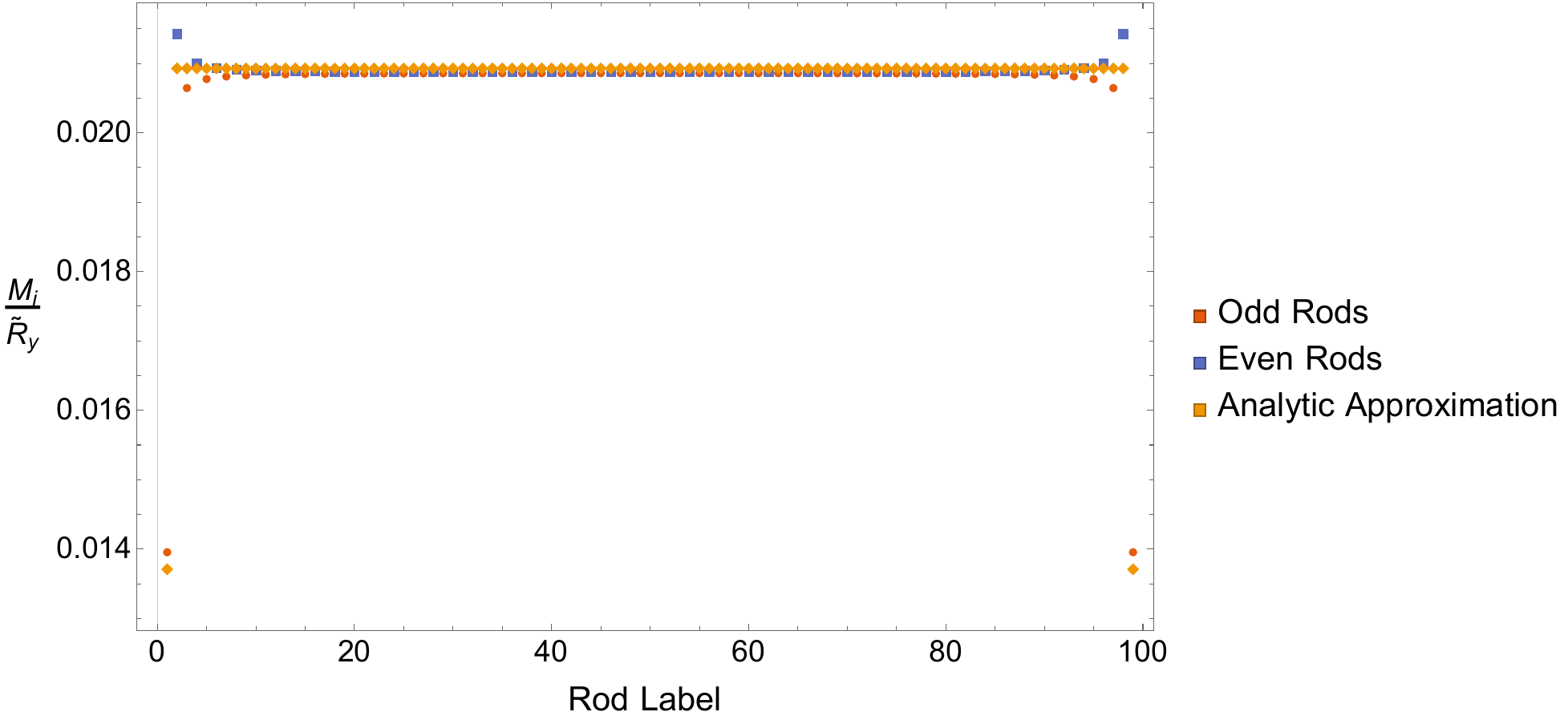}
\end{adjustwidth}
\caption{Plots of the rod lengths, $M_i$, as a function of the rod label, $i$, for a configuration of 99 bubbles. The ``odd and even rods'' dots are the numerical values while the ``analytic approximation'' ones correspond to our analytic computation.}
\label{fig:RodSizeApprox}
\end{figure}

We do the same technique as above but with now $M_1 = M_{2N+1} = m_1$ and $M_i=m$. By taking the large $N$ expansion of the bubble equations we find
\begin{equation}
\begin{split}
&\frac{2 m_1 \sqrt{\Gamma\left(\frac{m_1}{2m}\right)\,\Gamma\left(\frac{1}{2}+\frac{m_1}{2m}\right)}}{\Gamma\left(1+\frac{m_1}{2m}\right)} \,N^\frac{3}{4}\=  \widetilde{R}_{y}\,,\\
& \frac{2^{j(1+2j)+\frac{1}{2}\left(\frac{m_1}{m}-1\right)}(2j)!\,m}{j!\,\pi^{j+\frac{1}{4}}} \left( \frac{\binom{2j+1}{j}}{(j+2)\,j!}\right)^2\,\frac{\Gamma\left(j+\frac{m_1+m}{2m}\right)}{\sqrt{\Gamma\left(2j+\frac{m_1}{m}\right)}}\,N^\frac{3}{4}\, \prod_{k=1}^{j-1} \binom{2k+4}{k+1}^{-2} \= \widetilde{R}_{y}, \label{eq:BElargeNApp}
\end{split}
\end{equation}
where $1\leq j\leq \left\lfloor\frac{N+1}{2}\right\rfloor$. Taking the large $N$ expansion for the middle rod $ j= \left\lfloor\frac{N+1}{2}\right\rfloor$ leads to
\begin{equation}
\frac{2^\frac{25}{12}\,e^{\frac{1}{4}}\,m}{\cA^3} \, N^\frac{3}{4} \= \widetilde{R}_{y}\, \quad \Rightarrow \quad m \=N^{-\frac{3}{4}} \,\frac{\cA^3}{2^\frac{25}{12}\,e^{\frac{1}{4}}}\, \widetilde{R}_{y} \,.
\end{equation}
We can now fix $m_1$ in order for the first equation in \eqref{eq:BElargeNApp} to be identical to the one above. We find
\begin{equation}
\frac{ m_1 \sqrt{\Gamma\left(\frac{m_1}{2m}\right)\,\Gamma\left(\frac{1}{2}+\frac{m_1}{2m}\right)}}{m\,\Gamma\left(1+\frac{m_1}{2m}\right)} \=\frac{2^\frac{13}{12}\,e^{\frac{1}{4}}\,m}{\cA^3} \quad \Rightarrow \quad m_1 \approx 0.655\,m\,.
\end{equation}
Obviously, all the other equations in \eqref{eq:BElargeNApp} with different $j$ are not exactly the same. However, they are very close to each other, and should be made equal by considering small perturbations for each rod length $M_i = m + \cO(N^{-1})$. To illustrate this, we have plotted in Fig.\ref{fig:RodSizeApprox} the values obtained from the approximation for $N=49$ and compare it to the numerical data discussed before. Our analytic procedure is already very accurate even if $N$ is not so large.

\subsection{Analysis when $\widetilde{R}_{y_1} \neq \widetilde{R}_{y_2}$}
\label{app:BEsolutionsdifferentRy}

We have also been interested in solving the bubble equations when $\widetilde{R}_{y_1} \neq \widetilde{R}_{y_2}$. Since $\widetilde{R}_{y_1}$ is related to the length of the odd rods (the species-1 bubbles) and $\widetilde{R}_{y_2}$ is related to the length of the even rods (the species-2 bubbles), we expect that the ratio $\widetilde{R}_{y_1} / \widetilde{R}_{y_2}$ dial the density of species-1 bubbles with respect to the species-2 bubbles in the configuration. The question is how the rod lengths, $M_i$, depend on this ratio. 

We have solved the bubble equations numerically when  $\widetilde{R}_{y_1} \neq \widetilde{R}_{y_2}$ \eqref{eq:RegRyconnected}. We have found similar behaviors as in Fig.\ref{fig:RodSizeNum}. We found that all the even rods have mostly the same length which scales like $N$ to some negative power in between $-1$ and $0$ depending on the value of $\widetilde{R}_{y_2}/\widetilde{R}_{y_1}$, and the first and last species-2 bubbles in the chain can have a different length from the other if $\widetilde{R}_{y_2}\gg\widetilde{R}_{y_1}$. Moreover, the odd rods share the same property by reversing the role of $\widetilde{R}_{y_1}$ and $\widetilde{R}_{y_2}$. However, instead of having the lengths of the even and odd rods to be mostly the same as in Fig.\ref{fig:RodSizeApprox}, their mean values have a non-trivial ratio which is a function of $\frac{M_{2i+1}}{M_{2i}} \sim h\left(N,\frac{\widetilde{R}_{y_2}}{\widetilde{R}_{y_1}}\right)$ with $h\left(N,\frac{\widetilde{R}_{y_2}}{\widetilde{R}_{y_1}}\right) \sim 1$. Moreover, the difference between the lengths of the rods at the periphery and the lengths of the middle rods can vary significantly according to $\widetilde{R}_{y_2}/\widetilde{R}_{y_1}$ and can be large for instance when $\widetilde{R}_{y_1} /\widetilde{R}_{y_2} \gg1$ or $\ll1$.

A better understanding of the bubble equations in this regime will be required to have an exact description of Weyl stars with $\widetilde{R}_{y_1} \neq \widetilde{R}_{y_2}$ and the generic bag spacetimes they resolve as described in section \ref{sec:bagspacetimes}.

\section{Approximation out of the star surface}
\label{App:RSanalysis}

In this section we detail the procedure used in section \ref{sec:outofstar} where the Riemann sum approximation was applied to simplify the expressions for the warp factors and gauge potentials, \eqref{eq:WarpFactorsVacMultiple} and \eqref{eq:WarpFactorsVacMultipleCharged}, off the surface of the star $r>2M$. The method is the same for each function, so we will do a careful analysis only for the warp factor $U_1$,
\begin{equation}
U_1 \= \prod_{i=1}^{N+1} \left(1- \frac{2M_{2i-1}}{r^{(2i-1)}_+ +r^{(2i-1)}_-+M_{2i-1}} \right)\,.
\end{equation}
The Riemann sum approximation states that, for continuous functions $f$ and $g$, 
\begin{equation}
\sum_{i=1}^N f(x_i)\,\delta x \,\sim\, \int_{x_1}^{x_N} f(x) \,dx \quad \Leftrightarrow \quad \prod_{i=1}^N g(x_i) \,\sim \, \exp \left[\frac{1}{\delta x}\,\int_{x_1}^{x_N} \log g(x) \,dx \right] 
\label{eq:IntApproxApp}
\end{equation}
if $f=\log g$ is slowly varying in between each segment. 

\subsection{Regime of validity}

One can estimate the error using the generic inequality when $f$ is derivable
\begin{equation}
\left| \int_{x_1}^{x_N} f(x) \,dx - \sum_{i=1}^N f(x_i)\,\delta x \right| \leq \frac{(x_N-x_1) \delta x}{2}\,\left| \underset{x\in [x_1,x_N]}{\text{max}} (f'(x)) \right| \,.
\end{equation}
Note that this bound can give a poor estimation of what the real error is. Indeed the bound is saturated when $f$ is a linear function, but in our case $f$ will be highly picked on small localized regions, which means that $\underset{x\in [x_1,x_N]}{\text{max}} (f'(x))$ can be large in tiny regions where the integral is almost zero anyway. However, the bound still gives a sufficient condition to satisfy. If we find the domain of $(r,\theta)$ where 
\begin{equation}
 \frac{(x_N-x_1) \delta x}{2}\,\left| \underset{x\in [x_1,x_N]}{\text{max}} (f'(x)) \right| \ll 1 \,,\qquad \text{when } N\gg 1\,,
 \label{eq;conditionRSapp}
\end{equation} 
we are guaranteed that the Riemann sum approximation will give very accurate results in this region. Because the bound is not really accurate for our functions, it is most likely that the region found will be smaller than the real region of validity.

To simplify $U_1$, we define $x_i \= \frac{i}{N^\frac{3}{4}} -\frac{ N^\frac{1}{4}}{2}$, which implies $\delta x \= N^{-\frac{3}{4}}$, $x_1 \sim -\frac{N^\frac{1}{4}}{2}$ and $x_N = \frac{N^\frac{1}{4}}{2}$, and the function $g$ is given by
\begin{equation}
g(x) \equi 1 \- \frac{2}{1 \+ R(x) \+ R\left(x - \frac{1}{2N^\frac{3}{4}}\right)}
\end{equation}
where we have defined
\begin{equation}
R(x) \equi \sqrt{\left(\frac{2N+1}{2} \left( \frac{r}{M}-1\right) \cos \theta -2 N^\frac{3}{4}\,x \right)^2 + (2N+1)^2 \frac{r}{2M} \left( \frac{r}{2M}-1 \right) \sin^2 \theta}\,.
\label{eq:R(x)App}
\end{equation}
Therefore we have\footnote{We have considered that all $M_i$ are equal to $M_i = \kappa N^{-\frac{3}{4}} \widetilde{R}_y$ \eqref{eq:SolutionBE}. The fact that $M_1$ and $M_{2N+1}$ are slightly different is subleading in the large $N$ limit. Therefore, we took $z_i^\pm = N^{-\frac{3}{4}} \,\left(i-1-N \pm \frac{1}{2} \right) \kappa  \widetilde{R}_y$ \eqref{eq:zintouchingVac}.}
\begin{equation}
U_1 \= \prod_{i=1}^N g(x_i)\,.
\end{equation}
We can use the Riemann sum approximation \eqref{eq:IntApproxApp} if \eqref{eq;conditionRSapp} is satisfied with $f=\log g$. The expression of $f'(x)$ is rather complicated in the interval $x \in [- \frac{N^\frac{1}{4}}{2},\frac{N^\frac{1}{4}}{2}]$. We first rescale $x$ to be independent of $N$ such that 
\begin{equation}
w \= \frac{x}{N^\frac{1}{4}} \,,\qquad w \in [- \frac{1}{2},\frac{1}{2} ]\,.
\end{equation} 
The large $N$ expansion gives 
\begin{equation}
 \frac{(x_N-x_1) \delta x}{2}\,f'(x) \= \frac{1}{\sqrt{N}}\,f'(x) \= \frac{1}{N^\frac{1}{4}} \sum_{j=1}^\infty \frac{f_k(w,r,\theta)}{\left(h(w,r,\theta) N\right)^{k+1/2} }\,,
 \label{eq;conditionRSapp2}
\end{equation} 
where $f_k$ are complicated functions but finite and well-defined, while $h$ is given by
\begin{equation}
\begin{split}
h(w,r,\theta) \equi& \left( \frac{r}{2M}-\frac{1}{2} \left( 1+2 w \cos \theta  + \sqrt{1-4 w^2} \sin \theta\right)  \right) \\
&\times \left( \frac{r}{2M} -\frac{1}{2} \left( 1+2 w \cos \theta  \- \sqrt{1-4 w^2} \sin \theta\right)  \right)\,.
\end{split}
\end{equation}
We therefore needs to find the region of $(r,\theta)$ where $h(w,r,\theta) \gtrsim \frac{1}{N}$ for any $w \in [- \frac{1}{2},\frac{1}{2} ] $. This will imply, from \eqref{eq;conditionRSapp2}, that $N^{-\frac{1}{2}}  \underset{x\in [x_1,x_N]}{\text{max}} (f'(x)) \lesssim N^{-\frac{1}{4}}$ which guarantees the accuracy of the Riemann sum approximation \eqref{eq;conditionRSapp}.

We remind that the physical regime of $(r,\theta)$ is $r\geq 2M$ and $0 \leq \theta\leq \pi$. With a careful look to the expression of $h$, we can check that the $(w,\theta)$ parts can be close to $1$, and therefore $h$ can be close to zero, only at a specific point $w$ for fixed $\theta$ given by
\begin{equation}
w \= \frac{\cos \theta}{2} \quad \Rightarrow \quad h_\text{min} (r,\theta) \=\left( \frac{r}{2M}-1 \right) \left(\frac{r}{2M}-\cos^2 \theta \right)\,.
\end{equation}
Thus, imposing $h(w,r,\theta) \gtrsim \frac{1}{N}$ requires
\begin{equation}
\begin{split}
& r \gtrsim 2M \left( 1 \+ \cO \left( \frac{1}{N} \right) \right) \,,\quad \text{if }\theta \neq 0,\pi\,,\\
& r \gtrsim 2M \left( 1 \+ \cO \left( \frac{1}{\sqrt{N}} \right)\right) \,,\quad \text{if }\theta = 0,\pi\,,
\end{split}
\label{eq:regimevalidityRS}
\end{equation}
In this regime we have, from \eqref{eq;conditionRSapp}, 
\begin{equation}
\left| \int_{x_1}^{x_N} \log g(x) \,dx - \sum_{i=1}^N \log g(x_i)\,\delta x \right| \lesssim \frac{1}{N^\frac{1}{4}} \,,
\end{equation}
which allows to take \eqref{eq:IntApproxApp}, 
\begin{equation}
U_1 \sim \exp \left[N^\frac{3}{4}\,\int_{-\frac{N^{1/4}}{2}}^{\frac{N^{1/4}}{2}} \log g(x) \,dx \right] \= \exp \left[N\,\,\int_{-\frac{1}{2}}^{\frac{1}{2}} \log g\left(N^\frac{1}{4}\,w\right) \,dw \right]\,.
\end{equation}
The regime of validity is therefore very close to the surface of the star $r=2M$, and the Riemann sum approximation will give a trustworthy description of the Weyl star until very close to its surface. Moreover, note that the description is less accurate around the pole of the surface than elsewhere.

\subsection{Applying the integral formula}

We first expand $R(x)$ at large $N$ \eqref{eq:R(x)App}
\begin{equation}
R\left(N^\frac{1}{4}\,w\right) \= \sqrt{2} \sqrt{h(w,r,\theta)}\,N \left( 1\+ \cO\left( \frac{1}{N\,h(w,r,\theta) } \right) \right)
\end{equation}
In the regime given by \eqref{eq:regimevalidityRS}, we can then take $R\left(N^\frac{1}{4}\,w\right) \sim \sqrt{2} \sqrt{h(w,r,\theta)}\,N$. Inserting these expansions into $\log g\left(N^\frac{1}{4}\,w\right)$, we find 
\begin{equation}
\log g\left(N^\frac{1}{4}\,w\right) \sim - \frac{\sqrt{2}}{N\,\sqrt{h(w,r,\theta)}}\,,
\end{equation}
which is hopefully derivable and gives
\begin{equation}
N\,\,\int_{-\frac{1}{2}}^{\frac{1}{2}} \log g\left(N^\frac{1}{4}\,w\right) \,dw \sim \frac{1}{2} \, \log \left[ 1 - \frac{2M}{r} \right]\,.
\end{equation}
As a final result, the Riemann sum approximation leads to
\begin{equation}
U_1 \sim \sqrt{1- \frac{2M}{r}}\,,
\end{equation}
where $r$ and $\theta$ are in the regime of validity \eqref{eq:regimevalidityRS}. One can apply the same strategy to the other warp factors and gauge potentials \eqref{eq:WarpFactorsVacMultiple} and \eqref{eq:WarpFactorsVacMultipleCharged}. The analysis for $U_2$ will be very similar, and $H_1$ and $H_0$ will lead to
\begin{equation}
H_0 \sim \frac{3 M}{2 \sinh b_0} \, \cos \theta \,,\qquad H_1 \sim \frac{M}{2 \sinh b_1} \, \cos \theta \,,
\end{equation}
in the same regime of $(r,\theta)$ as above. The most difficult part is to simplify $e^{2\nu}$, which is a double product and therefore leads to a double integral. After some effort, we found
\begin{equation}
e^{2 \nu} \sim \left(\frac{r(r-2M)}{(r-M)^2-M^2 \cos^2\theta} \right)^\frac{3}{4}\,,
\end{equation}
in the same regime of $(r,\theta)$.



\bibliographystyle{utphys}      

\bibliography{microstates}       


\end{document}